\documentclass[]{aa}
\usepackage[dvips]{graphicx}
\usepackage{txfonts}
\begin{document}
\title{Tests of stellar model atmospheres by optical interferometry IV}
\subtitle{VINCI interferometry and UVES spectroscopy of Menkar
\thanks{Based on observations collected at the European Southern Observatory, 
Chile, under program number 71.D-0370(A,B). 
}}
\author{
M.~Wittkowski\inst{1} \and
J.~P.~Aufdenberg\inst{2} \and
T.~Driebe\inst{3} \and
V.~Roccatagliata\inst{4} \and
T.~Szeifert\inst{5} \and
B.~Wolff\inst{1}
}
\institute{
European Southern Observatory, Karl-Schwarzschild-Strasse 2, 
D-85748 Garching bei M\"unchen, Germany, \email{mwittkow@eso.org}
\and
National Optical Astronomy Observatory, 950 North Cherry Avenue, 
Tucson, AZ 85719, USA
\and
Max-Planck-Institut f\"ur Radioastronomie, Auf dem H\"ugel 69, 53121 Bonn,
Germany
\and
Max-Planck-Institut f\"ur Astronomie, K\"onigsstuhl 17,
69117 Heidelberg, Germany
\and
European Southern Observatory, Casilla 19001, Santiago 19, Chile
}
\titlerunning{Interferometry and spectroscopy of Menkar}
\date{Received \dots; accepted \dots}
\abstract
{}
{We present coordinated near-infrared $K$-band
interferometric and optical spectroscopic observations of 
the M\,1.5 giant \object{$\alpha$\,Cet} (Menkar) 
obtained with the instruments VINCI and UVES at the Paranal Observatory. 
Spherically 
symmetric \protect{\tt PHOENIX} stellar model atmospheres are constrained 
by comparison to our interferometric and spectroscopic
data, and high-precision fundamental parameters of Menkar are obtained.}
{Our high-precision VLTI/VINCI observations in the first and second lobes
of the visibility function directly probe the model-predicted strength of the 
limb darkening effect in the $K$-band and the stellar angular diameter.
The high spectral resolution of UVES of $R=80\,000-110\,000$ allows us to 
confront in detail observed and model-predicted profiles of atomic lines 
and molecular bands.}
{We show that our derived \protect{\tt PHOENIX} model atmosphere for
Menkar is consistent with both the measured strength
of the limb-darkening in the near-infrared $K$-band and the 
profiles of spectral bands around selected atomic lines 
and TiO bandheads from 370\,nm to 1000\,nm. At the detailed level
of our high spectral resolution, however, noticeable
discrepancies between observed and synthetic spectra exist.
We obtain a high-precision
Rosseland angular diameter 
of $\Theta_\mathrm{Ross}$=12.20\,mas\,$\pm$\,0.04\,mas.
Together with the Hipparcos parallax of 14.82\,mas\,$\pm$\,0.83\,mas, it
corresponds to a Rosseland radius 
of $R_\mathrm{Ross}$=89\,$\pm$\,5\,R$_\odot$, 
and together with
the bolometric flux based on available spectrophotometry, to an 
effective temperature of $T_\mathrm{eff}$=3795\,K\,$\pm$\,70\,K. The 
luminosity based on these values 
is $L$=1460\,$L_\odot$\,$\pm$\,300\,$L_\odot$.
Relying on stellar evolutionary tracks, these values correspond 
to a mass $M=$\,2.3\,$M_\odot$\,$\pm$\,0.2\,$M_\odot$ and a surface
gravity $\log g$=0.9\,$\pm$\,0.1 (cgs). }
{Our approach illustrates the power of combining interferometry and 
high-resolution spectroscopy to constrain and calibrate stellar model 
atmospheres. The simultaneous agreement of the model atmosphere with 
our interferometric and spectroscopic data increases confidence in the 
reliability of the modelling of this star, while discrepancies
at the detailed level of the high resolution spectra can be used
to further improve the underlying model.}
{}
\maketitle
\section{Introduction}
\label{sec:intro}
Stellar atmosphere models predict in general the spectrum emerging
from every point of a stellar disc. These models are often based 
on assumptions of a stationary 1-dimensional plane-parallel 
atmospheric structure (e.g., {\tt ATLAS\,9} models, Kurucz \cite{kurucz93}), 
or a stationary spherically symmetric structure (e.g., {\tt PHOENIX} models, 
Hauschildt \& Baron \cite{hauschildt99a}; 
Hauschildt et al. \cite{hauschildt99b}). 

Confrontation of model atmospheres with observations is often
performed by spatial integration of the model radiation field over the 
whole stellar disc, spectral integration
to broadband colours and/or spectrograms of different spectral
resolution, and subsequent comparison to observed broadband photometry,
spectro-photometry, or (high-resolution) spectroscopy.
For instance, Tripicchio et al. (\cite{tripicchio97,tripicchio99})
measured the D resonance lines of neutral sodium and the KI resonance line 
at 7699\,\AA\ to test {\tt MARCS} (Gustafsson et al. \cite{gustafsson75};
Plez et al. \cite{plez92,plez93}) model atmospheres that include 
photospheric and chromospheric effects. 
Valenti et al. (\cite{valenti98}) performed a least-squares fit of synthetic
spectra to echelle spectroscopy of an M\,3.5
dwarf. They successfully constrained the model atmosphere by
using the 7087\,\AA\ TiO bandhead and five
atomic Fe\,I and Ti\,I lines between 8674\,\AA\ and 8692\,\AA.
Decin et al. (\cite{decin03}) compared infrared spectroscopy obtained with 
ISO-SWS with {\tt MARCS} stellar model 
atmosphere predictions to estimate fundamental stellar parameters.

Optical interferometry provides a further and more direct test of 
stellar atmosphere models by resolving the stellar disc and measuring the 
centre-to-limb intensity variation (CLV) across the stellar disc.
Interferometric measurements of stellar CLVs include, for 
instance, those by Hanbury Brown et al. (\cite{hanbury74}), 
Haniff et al. (\cite{haniff95}), Quirrenbach et al. (\cite{quirrenbach96}), 
Burns et al. (\cite{burns97}), Hajian et al. (\cite{hajian98}), 
Wittkowski et al. (\cite{wittkowski01,wittkowski04,wittkowski06}), 
Perrin et al. (\cite{perrin04a,perrin04b}), 
Woodruff et al. (\cite{woodruff04}), Fedele et al. (\cite{fedele05}), and
Aufdenberg et al. (\cite{aufdenberg05}).
The combination of interferometry and high-resolution spectroscopy
to constrain stellar model atmospheres, as suggested and presented,
for instance, by Aufdenberg et al. (\cite{aufdenberg03}) and
van Belle et al. (\cite{vanbelle03}), has 
so far been surprisingly rare, although this approach can provide a 
very strong test of theoretical models.

Here, we present coordinated near-infrared $K$-band 
interferometry obtained with the VLTI instrument 
VINCI and high-resolution ($R$ up to 110\,000)
optical spectroscopy obtained with the VLT echelle spectrograph UVES  
of the M\,1.5 giant $\alpha$~Cet 
(Menkar, HD\,18884, HR\,911, HIP\,14135).
We demonstrate the advantage of combined interferometry and
high-resolution spectroscopy to test and constrain {\tt PHOENIX} 
model atmospheres (Hauschildt \& Baron \cite{hauschildt99a};
Hauschildt et al. \cite{hauschildt99b}). We derive a set of
high-precision fundamental parameters of $\alpha$~Cet.

$\alpha$\,Cet is a red giant with spectral type M\,1.5\,III
(Morgan \& Keenan \cite{morgan73}), a parallax
of $\pi$=14.82\,mas\,$\pm$\,0.83\,mas, a $V$ magnitude of 2.54
(both from Perryman \& ESA \cite{perryman97}), and a small variability 
of $\Delta V$=0.09 (Kholopov et al. \cite{kholopov99}).
Decin et al. (\cite{decin03}) derived an effective temperature 
of $T_\mathrm{eff}$=3740\,$\pm$\,140 K, a surface gravity
of $\log g$=0.95\,$\pm$\,0.25, a mass of $M/M_\odot$=2.69\,$\pm\,$1.61,
a [Fe/H]$=0.00\ \pm 0.30$,
a luminosity of $L$=1455\,$\pm$\,328\,$L_\odot$, and a (limb-darkened) angular 
diameter $\Theta$ of 12.52\,$\pm$\,0.79\,mas, based on a 
comparison of ISO SWS data to {\tt MARCS} models.
Alonso et al. (\cite{alonso00}) derived a (limb-darkened) angular 
diameter of 12.59\,$\pm$\,0.36\,mas and an effective temperature 
of 3704\,$\pm$\,39\,K by means of the infrared flux method (IRFM).
Earlier interferometric diameter measurements include those by
Dyck et al. (\cite{dyck98}: $\Theta_\mathrm{UD}$
at $K$-band 11.6\,$\pm$\,0.4\,mas; bolometric flux 
1.05$\pm$0.16\,10$^{-12}$\,W\,cm$^{-2}$),
Quirrenbach et al. (\cite{quirrenbach93}: $\Theta_\mathrm{UD}$
at 712\,nm, TiO band, 11.95\,$\pm$0.23\,mas; $\Theta_\mathrm{UD}$ at 754\,nm,
continuum, 11.66\,$\pm$\,0.22\,mas), 
Mozurkewich et al. (\cite{mozurkewich91,mozurkewich03}: 
$\Theta_\mathrm{UD}$ at 800\,nm 12.269$\pm$0.237\,mas; $\Theta_\mathrm{UD}$ 
at 550\,nm 11.473$\pm$0.251\,mas; $\Theta_\mathrm{UD}$ at 451\,nm 
11.325$\pm$0.410\,mas).
\section{VINCI observations and data reduction}
\label{sec:vinci}
\begin{table}
\centering
\caption{Properties of the used interferometric calibration stars.
The references are (D) calibration by Dyck et al. (\cite{dyck96}), 
(B) Bord\'e et al. (\cite{borde02}), and (K) 
Kervella et al. (\cite{kervella03b}).}
\begin{tabular}{llllr}
\hline\hline
Calibrator &$\Theta_\mathrm{UD}$ & Ref. & Sp. Type & $T_\mathrm{eff}$\\ \hline
 17\,Mon         &  2.59$\pm0.26$ & D  & K4III    & 4090\\
 18\,Mon         &  1.86$\pm0.02$ & B  & K0IIIa   & 4656\\
 19\,Ari         &  2.37$\pm0.03$ & B  & M0III    & 3690\\
 28\,Mon         &  3.14$\pm0.30$ & D  & K4III    & 4090\\
 29\,Ori         &  1.92$\pm0.19$ & D  & G8III    & 4670\\
 30\,Gem         &  2.02$\pm0.03$ & B  & K1III    & 4581\\
 31\,Leo         &  3.22$\pm0.05$ & B  & K4III    & 4202\\
 31\,Ori         &  3.56$\pm0.06$ & B  & K5III    & 4046\\
 6\,Leo          &  2.10$\pm0.03$ & B  & K2.5IIIb & 4318\\
 $\beta$\,Cet    &  5.18$\pm0.06$ & B  & K0III    & 4656\\
 $\chi$\,Phe     &  2.69$\pm0.03$ & B  & K5III    & 4046\\
 HR\,1663        &  2.69$\pm0.27$ & D  & K5III    & 3920\\
 $\eta$\,Cet     &  3.35$\pm0.04$ & B  & K1.5III  & 4380\\
 $\eta$\,Eri     &  2.64$\pm0.20$ & D  & K1III    & 4510\\
 $\gamma^1$\,Cae &  2.31$\pm0.23$ & D  & K3III    & 4230\\
 $\alpha$\,Car   &  6.45$\pm0.60$ & D  & F0II     & 7200\\
 HR\,1799        &  2.04$\pm0.20$ & D  & K5III    & 3920\\
 HR\,2311        &  2.43$\pm0.04$ & B  & K5III    & 4046\\
 HR\,4546        &  2.53$\pm0.04$ & B  & K3III    & 4256\\
 HR\,6862        &  2.59$\pm0.05$ & B  & K4.5III  & 4097\\
 HR\,7092        &  2.81$\pm0.03$ & B  & M0III    & 3690\\
 $\iota$\,Cet    &  3.27$\pm0.04$ & B  & K1.5III  & 4508\\
 $\iota$\,Eri    &  2.12$\pm0.02$ & B  & K0III    & 4581\\
 $\iota$\,Hya    &  3.41$\pm0.05$ & B  & K2.5III  & 4318\\
 $\nu^2$\,CMa    &  2.38$\pm0.03$ & B  & K1III    & 4497\\
 $\alpha$\,CMa   &  5.93$\pm0.02$ & K  & A1V      & 9230\\
 $\theta$\,CMa   &  4.04$\pm0.40$ & D  & K4III    & 4099\\
 $\zeta$\,Hya    &  3.33$\pm0.30$ & D  & G9II-III & 4559\\ \hline
\end{tabular}
\label{tab:calibrators}
\end{table}
We obtained near-infrared $K$-band interferometric measurements 
of $\alpha$\,Cet in the first and second lobe of the visibility function  
using the ESO Very Large Telescope Interferometer (VLTI) equipped with the
VINCI instrument and the two VLTI siderostats. 
For a recent technical description of the VLTI, see for instance 
Glindemann et al. (\cite{glindemann03}); for the VINCI instrument, see
Kervella et al. (\cite{kervella03,kervella04}).

Between 2 October 2003 and 5 December 2003, 38 visibility measurements
were secured in the first lobe of the visibility function using the
baseline E0-G0 with an unprojected baseline length of 16\,m.
Further observations obtained during the same period (11 additional nights
during October to December) and using the same 16\,m baseline were not used
because a consistent interferometric transfer function could not be 
established with sufficient precision, partly owing to an insufficient
time coverage of calibration stars.
On 11 \& 12 January 2004, four visibility measurements were 
recorded in the second lobe of the visibility function employing the D0-H0
baseline with an unprojected length of 64\,m.
Since all our observations cover only a small range of projected
baseline angles on the sky (see Table~\ref{tab:visibility}),
they are not sensitive to possible asymmetries of the stellar disc.
Data of $\alpha$~Cet and of several interferometric calibration
stars were obtained as series of 500 interferograms with scan lengths of
250\,$\mu$m and a fringe frequency of 295\,Hz. Calibration star details, 
including the adopted angular diameters, are listed 
in Table~\ref{tab:calibrators}.
\begin{table}
\centering
\caption{Details of our VINCI observations of $\alpha$\,Cet 
(date and time of observation,
spatial frequency, azimuth angle of the projected baseline (E of N)),
together with the measured squared visibility amplitudes and their
errors. The last column denotes the number of successfully
processed interferograms for each series.}
\label{tab:visibility}
\begin{tabular}{lccccl}
\hline\hline
Date \& & Sp. freq  & az  & $V^2$  & $\sigma_{V^2}$ & \# \\
Time (UT) &[1/$^{\prime\prime}$]& [deg] &   &  &   \\\hline
2003-10\\
-03 4:04:48 & 22.79&  67.9&    8.550e-01&   1.090e-02&      488\\
-03 4:12:01 & 23.57&  68.6&    8.322e-01&   1.073e-02&      473\\
-03 4:44:19 & 26.86&  70.7&    7.782e-01&   1.002e-02&      486\\
-03 4:51:27 & 27.53&  71.1&    7.629e-01&   9.888e-03&      467\\
-03 5:25:02 & 30.37&  72.4&    7.150e-01&   9.299e-03&      490\\
-03 5:32:18 & 30.91&  72.5&    7.052e-01&   9.208e-03&      482\\
-20 6:23:10 & 35.47&  72.8&    6.202e-01&   2.874e-02&      445\\
-20 6:30:14 & 35.48&  72.7&    6.321e-01&   2.951e-02&      414\\
-21 5:58:25 & 35.27&  73.1&    6.128e-01&   1.737e-02&      455\\
-21 6:05:40 & 35.37&  73.0&    6.062e-01&   1.718e-02&      471\\
-22 3:29:06 & 26.81&  70.7&    7.951e-01&   1.044e-02&      442\\
-22 3:36:30 & 27.51&  71.1&    7.712e-01&   9.284e-03&      425\\
-22 4:15:05 & 30.73&  72.5&    7.169e-01&   1.117e-02&      408\\
-22 4:22:10 & 31.24&  72.7&    7.073e-01&   1.302e-02&      354\\
-24 5:34:53 & 35.04&  73.2&    6.278e-01&   2.122e-02&      479\\
-24 5:48:19 & 35.30&  73.1&    6.222e-01&   2.146e-02&      351\\
-29 5:47:00 & 35.47&  72.9&    6.420e-01&   1.237e-02&      348\\
-29 5:58:10 & 35.47&  72.7&    6.573e-01&   8.384e-03&     467\\[1ex]
2003-11\\
-27 3:14:20 & 34.87&  73.3&    6.435e-01&   6.270e-03&      492\\
-27 3:21:37 & 35.05&  73.2&    6.393e-01&   6.255e-03&      488\\[1ex]
2003-12\\
-04 1:37:03 & 31.51&  72.7&    7.067e-01&   7.754e-03&      488\\
-04 1:44:08 & 31.98&  72.9&    7.033e-01&   7.740e-03&      479\\
-04 2:17:53 & 33.82&  73.2&    6.689e-01&   7.695e-03&      482\\
-04 2:24:54 & 34.12&  73.3&    6.560e-01&   7.413e-03&      476\\
-04 3:03:23 & 35.24&  73.1&    6.384e-01&   7.146e-03&      483\\
-04 3:10:23 & 35.34&  73.1&    6.367e-01&   7.200e-03&      475\\
-04 4:19:06 & 34.78&  71.5&    6.393e-01&   7.430e-03&      437\\
-04 4:26:25 & 34.56&  71.3&    6.437e-01&   9.154e-03&      388\\
-05 0:05:53 & 23.74&  68.7&    8.474e-01&   1.024e-02&      338\\
-05 0:13:52 & 24.59&  69.3&    8.037e-01&   2.597e-02&       93\\
-05 1:35:55 & 31.70&  72.8&    7.091e-01&   7.531e-03&      471\\
-05 1:43:08 & 32.16&  72.9&    6.969e-01&   7.743e-03&      467\\
-05 2:28:03 & 34.39&  73.3&    6.520e-01&   7.566e-03&      447\\
-05 2:35:39 & 34.65&  73.3&    6.548e-01&   9.280e-03&      358\\
-05 3:23:51 & 35.48&  72.8&    6.304e-01&   7.116e-03&      469\\
-05 3:49:52 & 35.32&  72.3&    6.388e-01&   7.630e-03&      424\\
-06 3:48:38 & 35.29&  72.2&    6.360e-01&   8.405e-03&      482\\
-06 3:55:49 & 35.16&  72.0&    6.376e-01&   8.533e-03&      483\\[1ex]
2004-01\\
-12 1:00:54 &141.86&  72.7&    1.281e-02&   6.655e-04&      275\\
-12 1:13:28 &141.57&  72.4&    1.300e-02&   6.133e-04&      299\\
-13 1:28:57 &140.38&  71.9&    1.324e-02&   6.934e-04&      275\\
-13 1:40:23 &139.25&  71.6&    1.365e-02&   6.689e-04&      270\\
\hline
\end{tabular}
\end{table}

Mean coherence factors were obtained for each series of interferograms 
using the VINCI data reduction software, version 3.0, as described by
Kervella et al. (\cite{kervella04}), employing the results based on the
wavelet transforms. Calibrated squared visibility values for $\alpha$~Cet
were obtained from the mean coherence factors as described in Wittkowski
et al. (\cite{wittkowski04}). A time kernel of 3 hours was used to convolve
the measured interferometric transfer function during each night.
The computed errors of the squared visibility amplitude take into account the 
scatter of the single scan's coherence factors, the errors of the 
diameters of the calibration stars,
and the variation of the interferometric transfer function over the night.
Observational details and the calibrated squared visibility values with their
errors are listed in Table~\ref{tab:visibility}.
The inverse of the mean wavenumber of the $\alpha$~Cet observations 
was 2.187\,$\mu$m.

As a first characterisation of the angular diameter of $\alpha$~Cet, 
we calculated best-fitting angular diameters $\Theta_\mathrm{UD}$
and $\Theta_\mathrm{FDD}$ for models of a 
uniform disc (UD: $I=1$ for $0 < \mu < 1$; else 0) and a fully darkened 
disc (FDD: $I=\mu$). Here, $\mu=\cos\theta$ is the cosine of the angle
between the line of sight and the normal of the surface element of the 
star. The angular
diameter ($\Theta_\mathrm{UD}$ or $\Theta_\mathrm{FDD}$) is the only
free parameter. Synthetic squared visibility amplitudes were calculated
using the broadband VINCI sensitivity function (cf. the description 
in Wittkowski et al. \cite{wittkowski04,wittkowski06}). We obtained 
$\Theta_\mathrm{UD}$=11.95\,mas\,$\pm$\,0.06\,mas 
with a reduced $\chi^2_\nu$(UD) value of 3.35, 
and $\Theta_\mathrm{FDD}$=13.32\,mas\,$\pm$\,0.12\,mas with 
$\chi^2_\nu$(FDD)=10.20.

Furthermore, we used an empirical parametrisation of the limb-darkened 
stellar intensity profile as $I(\mu)=\mu^\alpha$ 
(cf. Hestroffer et al. \cite{hestroffer97}), and simultaneously determined
the best-fitting values for the limb-darkened angular 
diameter $\Theta_\mathrm{LD}$ and the limb-darkening 
parameter $\alpha$. We obtain 
$\Theta_\mathrm{LD}$=12.27\,mas\,$\pm$\,0.05\,mas
and $\alpha$=0.24\,$\pm$0.03 with $\chi^2_\nu$(LD)=1.01.
This parametrisation of a limb-darkened intensity profile closely
resembles the CLVs (centre-to-limb variations) that are obtained by 
stellar model atmospheres
based on plane-parallel geometry, where the atmosphere is optically
thick for any viewing angle, and the intensity steeply drops to zero
at the stellar limb. In this case, the Rosseland
angular diameter and our obtained limb-darkened angular diameter,
which corresponds to the 0\% intensity level,
can be considered equivalent (cf. the discussion in 
Wittkowski et al. \cite{wittkowski04}).

\begin{figure}
\centering
\resizebox{0.945\hsize}{!}{\includegraphics{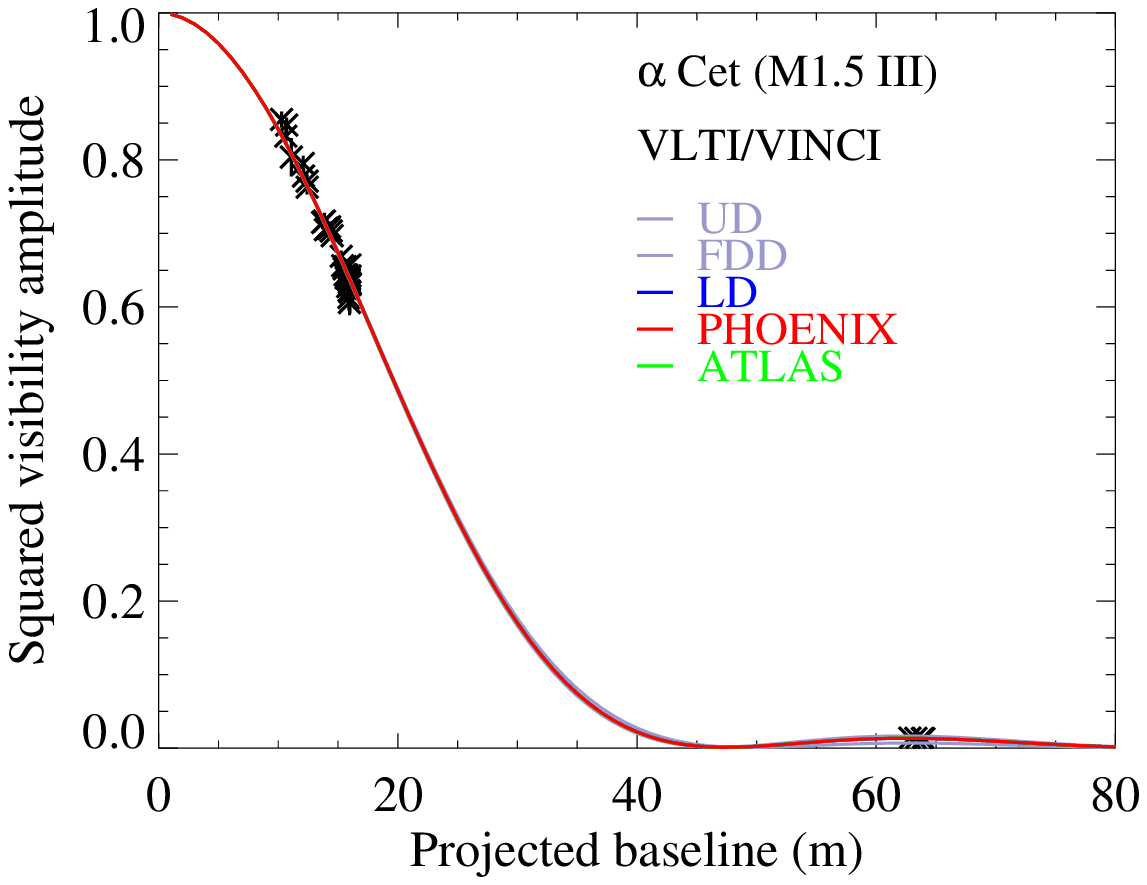}}
\resizebox{0.945\hsize}{!}{\includegraphics{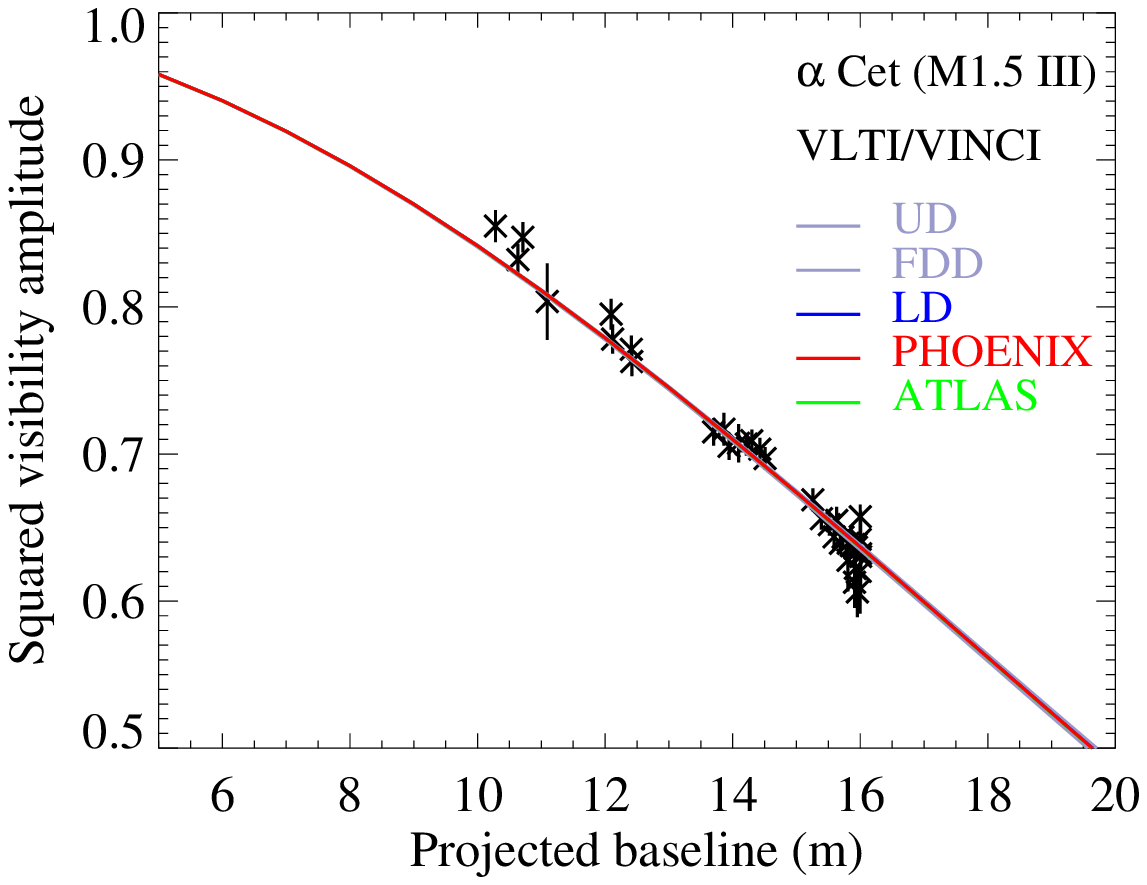}}
\resizebox{0.945\hsize}{!}{\includegraphics{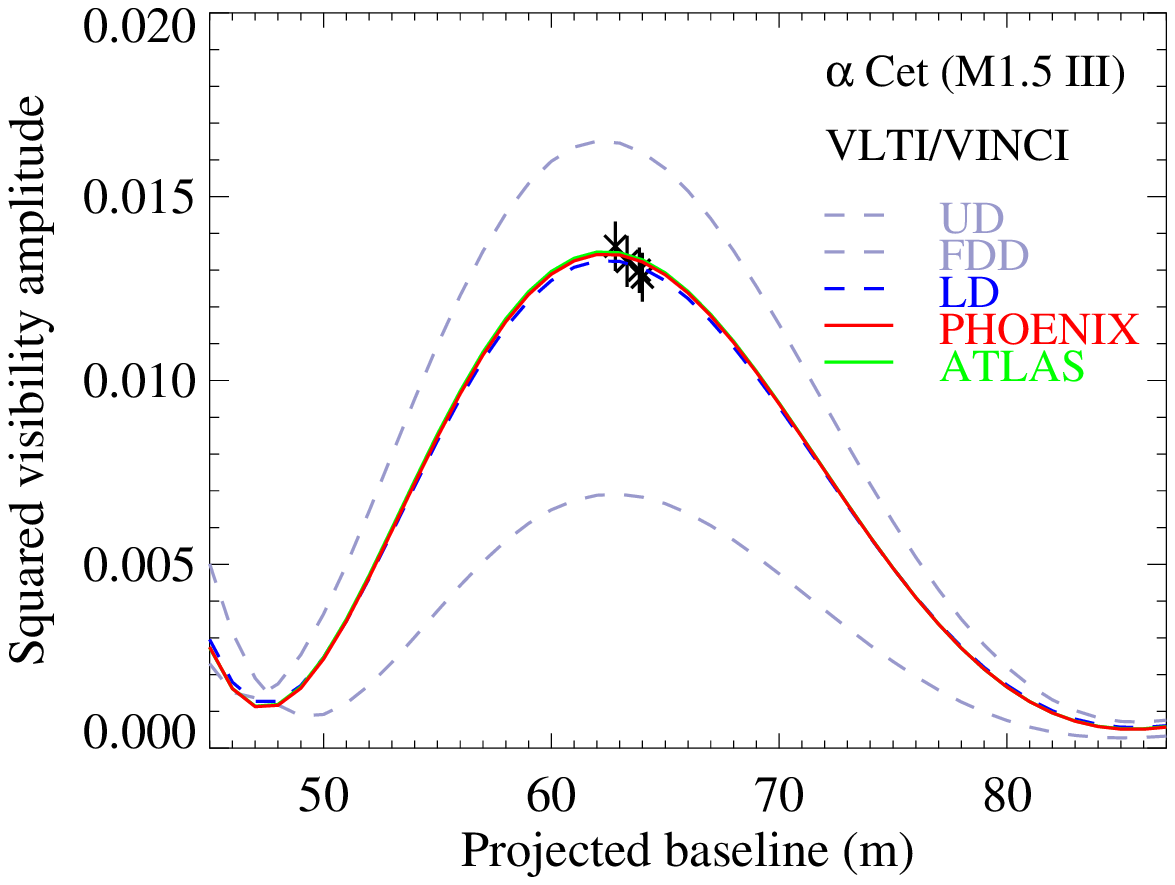}}
\caption{Squared visibility amplitudes and error bars of $\alpha$ Cet
obtained with VLTI/VINCI, together with best-fitting models of a
uniform disc (upper dashed light blue line), a fully darkened disc 
(lower dashed light blue line),
a parametrisation $I=\mu^{\alpha}$ with $\alpha$=0.24 (dashed blue line) 
and of {\tt PHOENIX} and {\tt ATLAS\,9} model atmosphere predictions (solid 
red and green lines). The {\tt PHOENIX} model shown 
has parameters $T_\mathrm{eff}$=3800\,K,
$\log g$=1.0, $M=2.3 M_\odot$; the {\tt ATLAS\,9} model 
$T_\mathrm{eff}$=3800\,K, $\log g$=1.0 (see text for more details).
The upper panel shows the full range of the 
visibility function, while the middle panel is an enlargement of the
obtained squared visibility values in the first lobe, and the bottom panel
shows an enlargement of the low squared visibility function in the 2nd
lobe. Our measurements are significantly different from uniform disc and
fully-darkened disc models, and well consistent with 
the LD ($I=\mu^{\alpha}$), 
{\tt PHOENIX}, and {\tt ATLAS\,9} models.}
\label{fig:vinciplot}
\end{figure}

These results already show that our interferometric data cannot
be well described by uniform or fully darkened disc models, but
are well consistent with a limb-darkened intensity profile that has
a strength of the limb-darkening ($\alpha$=0.24) that is closer to 
that of a  uniform disc ($\alpha$=0) 
than a fully-darkened disc ($\alpha$=1) model. 
Figure~\ref{fig:vinciplot} shows our measured visibility
points together with the described UD, FDD, and LD models, as well
as with the predictions by {\tt PHOENIX} and {\tt ATLAS\,9}  
model atmospheres (derived and described below).
\section{UVES observations and data reduction}
\label{sec:uves}
\begin{table}
\centering
\caption{Overview of our UVES observations of $\alpha$\,Cet 
obtained on 11 August 2003 between 9:30h and 9:40h UT.
Shown are the central wavelength $\lambda_\mathrm{central}$ of each
of the UVES gratings and the recorded wavelength 
ranges $\lambda_\mathrm{min}-\lambda_\mathrm{max}$. The spectra of the
red arm of UVES were recorded using two different detectors, and are
thus split into the two wavelength ranges. Also given
are, for each grating, the spectral resolution $R=\lambda/\Delta\lambda$,
the number of individual exposures and their exposure times, as well
as the maximum signal-to-noise ratio $S/N_\mathrm{max}$ reached.}
\begin{tabular}{rrrrrr}
\hline\hline
$\lambda_\mathrm{central}$ & $\lambda_\mathrm{min}-\lambda_\mathrm{max}$ &
$R$ & \# of  & Exp. time of & $S/N_\mathrm{max}$\\
(nm)    & (nm) & & exp.  & each exp. (s)\\ \hline
346 & 305-387 & 80\,000&2 &120.0 & 500\\[1ex]
437 & 374-499 & 80\,000&2 &3.7 & 280\\[1ex]
580 & 477-577/& 110\,000&7 &1.5 & 800\\
    & 583-683 & &\\[1ex]
860 & 665-853/& 110\,000&2 &0.5 & 280\\ 
    & 867-1040& \\\hline
\end{tabular}
\label{tab:UVES_obs}
\end{table}
We obtained high-resolution spectroscopy of $\alpha$\,Cet with the echelle 
spectrograph UVES mounted on the UT2 (KUEYEN) telescope of the ESO VLT 
in service mode on 11 August 2003. We used the 
dichroic settings 346+580\,nm and 437+860\,nm
with slit widths of 0.4$^{\prime\prime}$ (blue arm) and 0.3$^{\prime\prime}$ 
(red arm). This corresponds to a
spectral resolution of 80\,000 in the blue and 110\,000 in the red, 
respectively.
The details of our UVES data with numbers of individual integrations 
for each central wavelength of the UVES grating and exposure times 
are listed in Table~\ref{tab:UVES_obs}.

For all the observations, the bias and inter-order background were
subtracted. The spectral orders were extracted, flat-fielded, 
and wavelength calibrated with recipes available from the
ESO UVES pipeline. The ``average extraction method'' of the 
pipeline has been used, which uses a uniform average of the pixel values 
across the slit. Statistical errors have been calculated from the 
variance obtained with this extraction method.
The extracted spectra have been flux-calibrated 
using averaged response curves as provided by the 
Quality Control Group of ESO
Garching. This calibration provides a relative 
flux calibration with an accuracy of $\sim$10-20\% within each spectrum. 
Absolute flux values for our UVES spectrum of $\alpha$\,Cet have not 
been obtained.
Finally, multiple exposures of each setting have been averaged.
The resulting signal-to-noise ratios $S/N$ vary across the spectra,
and the maximum $S/N$ values reached for each grating are listed
in Table \ref{tab:UVES_obs}.
The correction to heliocentric velocity has been 
determined by the pipeline analysis to $v_\mathrm{helio}$=-28.7\,km/sec,
so that the relation between arriving wavelength from the star $\lambda_0$
and observed wavelength $\lambda_\mathrm{obs}$ 
is $\lambda_\mathrm{obs}=\lambda_0(1+v_\mathrm{helio}/c)$.
\section{Atmosphere models for $\alpha$\,Cet}
We use new, fully line-blanketed spherical, 
hydrostatic atmosphere models with solar photospheric 
abundances (Grevesse \& Noels \cite{grevesse93}) obtained
with version 13 of the {\tt PHOENIX} 
code (Hauschildt \& Baron \cite{hauschildt99a}; 
Hauschildt et al. \cite{hauschildt99b}).
Feast et al. (\cite{feast90}) discussed that solar metallicity
is appropriate for giants in the solar neighbourhood and found that 
the (rms) scatter of the metallicity of local giants is less than 0.08.
This is consistent with the [Fe/H]$=0.00\ \pm\ 0.30$ derived by
Decin et al. (\cite{decin03}) for $\alpha$\,Cet.
The microturbulence for all our new models is 2\,km\,s$^{-1}$.
The three most important input parameters for our spherical model
are the effective temperature, the surface gravity, and the stellar mass.
For each model used, we tabulate the flux integrated over the stellar disc
from 300\,nm to 1050\,nm in steps of 0.001 nm (for comparison to our high 
spectral resolution UVES data).
Furthermore, we tabulate monochromatic intensity profiles 
at 64 viewing angles for wavelengths from 1.8\,$\mu$m to 2.5\,$\mu$m in 
steps of 0.5\,nm (for comparison to our VINCI interferometric data). 

For comparison, we use intensity profiles predicted by standard
plane-parallel hydrostatic ATLAS\,9 model atmospheres from the Kurucz
CD-ROMs (Kurucz \cite{kurucz93}), as in 
Wittkowski et al. (\cite{wittkowski01,wittkowski04,wittkowski06}), as well. 
The Kurucz CD-ROMs include tabulated 
monochromatic intensity profiles for 17 angles in 1221 frequency intervals 
ranging from 9.09\,nm to 160\,000\,nm. In the range of the VINCI near-infrared
$K$-band filter, the frequencies are sampled in steps 
corresponding to 10\,nm. These data values are available as grids of 
effective temperature and surface gravity (the mass is not an input
parameter for a plane-parallel model), and for different chemical
abundances and microturbulent velocities. For comparison, we use 
the grid with solar chemical abundances and a microturbulence 
of 2\,km\,s$^{-1}$.

For a further description of the use of the {\tt PHOENIX}
and {\tt ATLAS} 
models for comparison to interferometric data obtained with VINCI, 
for the calculation of synthetic visibility values, and for effects
of plane-parallel and spherical geometries on the obtained angular
diameter, we refer to 
Wittkowski et al. (\cite{wittkowski04,wittkowski06}).
\subsection{The bolometric flux of $\alpha$~Cet}
We have constructed a composite spectral energy distribution for
$\alpha$~Cet by combining absolute spectrophotometry in the optical
(Glushneva et al. \cite{glushneva98a,glushneva98b}) and infrared
(Cohen et al. \cite{cohen96}) to estimate the bolometric
energy flux, $f_\mathrm{bol}$, at the Earth (see Fig.~\ref{fig:sed}).  
Note that an absolute flux calibration of our UVES echelle spectrum has not been obtained (see Sect.~\ref{sec:uves}),
so that we rely on spectrophotometry from other sources to derive
the bolometric flux of $\alpha$\,Cet.
We directly integrate
the tabulated irradiance from 322.5 nm to 1037.5 nm (5 nm resolution)
and 1.26\,$\mu$m to 35\,$\mu$m (0.02\,$\mu$m to 0.18\,$\mu$m
resolution), using a five-point Newton-Cotes integration formula.
Between 1.04\,$\mu$m and 1.26\,$\mu$m, the irradiance at 20 nm
intervals is estimated by linearly interpolating, in $\log\lambda -
\log F_{\lambda}$ space, between the end points of the two data sets.
We assume all the irradiance data are uncertain at the level of 5\%. This
procedure results in $f_\mathrm{bol}$=(1.01 $\pm$
0.05)\,10$^{-12}$\,W\,cm$^{-2}$.  At a distance of 67 pc, the
interstellar foreground extinction in the direction of $\alpha$~Cet is
likely to be quite low.  HD\,18883 (B7 V), at a distance of 134 pc, 
has a measured colour excess, $E(B-V)$=0.03, from space-based UV
photometry (Savage et al. \cite{savage85}) and is 0.3$^\circ$ from
$\alpha$~Cet.  Therefore, adopting $E(B-V)$=0.015 and an average $R_V$=3.1
Galactic reddening curve (Cardelli et al. \cite{cardelli89})
as a best estimate for the extinction towards $\alpha$~Cet, 
the bolometric flux becomes $f_\mathrm{bol}$=(1.03
$\pm$ 0.07)\,10$^{-12}$\,W\,cm$^{-2}$, including the uncertainty in the
extinction.  This value is well consistent with the estimate by Dyck et
al. (\cite{dyck98}) of (1.05$\pm$0.16)\,10$^{-12}$\,W\,cm$^{-2}$ based on
broadband photometry alone.

\subsection{Comparison of {\tt PHOENIX} model predictions to available
spectrophotometry}
\begin{figure*}
\centering
\resizebox{12cm}{!}{\includegraphics[angle=90]{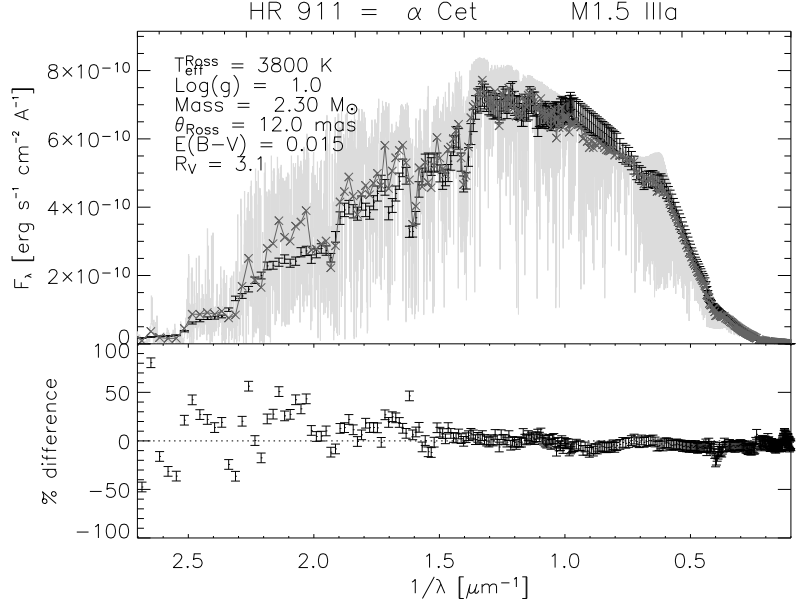}}
\caption{(Top) The observed spectral energy distribution of
$\alpha$~Cet (shown as error bars) from 
Glushneva et al. (\cite{glushneva98a,glushneva98b}) and Cohen et
al. (\cite{cohen96}) compared with synthetic spectrophotometry 
('{\tt x}' symbols)
derived from the {\tt PHOENIX} model (shown in grey at
high-resolution) with the mean parameters listed in the text.  
(Bottom) The percentage difference between the observed and 
synthetic spectrophotometry in each wavelength bin.}
\label{fig:sed}
\end{figure*}
We have compared synthetic spectral energy distributions (SEDs)
from {\tt PHOENIX} models 
with the absolute spectrophotometry used to derive the bolometric flux
above. 
The goodness of the SED
fit is largely insensitive to specific values of the surface gravity
and mass in the model grid.  The least-squares fits for the grid and
application of the F-test provide the following mean values and
1\,$\sigma$ uncertainties: $T_\mathrm{eff}$=3800\,$\pm$\,100\,K for
$\Theta_\mathrm{Ross}$=12.0\,$\pm$\,0.7\,mas.
These uncertainties are
correlated, so that low $T_\mathrm{eff}$ values correspond to 
high $\Theta_\mathrm{Ross}$ values and vice versa. 
Figure~\ref{fig:sed} shows the spectrophotometric data compared to the
synthetic SED based on the model with mean parameters mentioned above.
Differences between observed and synthetic SED are largest at
optical ($\lambda < 0.7$\,nm) wavelengths.
With the Hipparcos parallax, the
angular diameter of $\Theta_\mathrm{Ross}$=12.0\,$\pm$\,0.7\,mas
corresponds to a Rosseland linear radius of
$R_\mathrm{Ross}$=87\,$\pm$\,10\,$R_{\odot}$ and luminosity
$\log(L)$=3.16\,$\pm$\,0.18\,$L_{\odot}$.  
Relying on the evolutionary
tracks from Girardi et al. (\cite{girardi00}), the position of the
star in the theoretical HR diagram results in a mass
$M/M_\odot$=2.3\,$\pm$\,0.5 
(see Fig.~1 in Wittkowski et al. \cite{wittkowski04}).
From the mass and the Rosseland radius, we
derive a surface gravity of $\log g$=0.9\,$\pm$\,0.2.
\subsection{Comparison of model predictions to our VINCI data}
We derive best-fitting limb-darkened (0\% intensity) 
diameters $\Theta_\mathrm{LD}$ to our VINCI data for different
{\tt PHOENIX} model atmospheres with parameters $T_\mathrm{eff}$, 
$\log g$, and $M$ that lie within ranges consistent with the estimates
above. The procedure used is the same as described 
in Wittkowski et al. (\cite{wittkowski04,wittkowski06}). 
Table~\ref{tab:vinciresults} lists the considered model parameters and
the resulting best-fitting $\Theta_\mathrm{LD}$, $\Theta_\mathrm{Ross}$, 
and reduced $\chi^2_\nu$ values for each model. 
The ratio between the outermost model layer corresponding 
to $\Theta_\mathrm{LD}$ and the $\tau_\mathrm{Ross}$=1 layer corresponding
to $\Theta_\mathrm{Ross}$ is read from the {\tt PHOENIX} model 
atmospheric structure
as described in Wittkowski et al. (\cite{wittkowski04}).
The precision of the best-fitting diameter values for a given 
model atmosphere 
listed in Table~\ref{tab:vinciresults} is relatively high (0.4\%). 
We have shown by a study of night-to-night variations that 
such high-precision diameter values based on VINCI data are generally 
reliable (Wittkowski et al. \cite{wittkowski06}). Also, the different
considered {\tt PHOENIX} models with variations of $T_\mathrm{eff}$,
$\log g$, and $M$ result in consistent Rosseland angular diameters,
so that no uncertainty due to the choice of model parameters needs
to be added.

For comparison, we derive the best-fitting limb-darkened diameter
based on the corresponding plane-parallel {\tt ATLAS\,9} model as well.  
For plane-parallel geometry, $\Theta_\mathrm{LD}$ and
$\Theta_\mathrm{Ross}$ can be considered equivalent 
(Wittkowski et al. \cite{wittkowski04}). However, the definition
of the Rosseland diameter is more precise for models based on 
spherical geometry (cf. Wittkowski et al. \cite{wittkowski06}).
Here, the resulting Rosseland angular diameters based on the {\tt PHOENIX}
and {\tt ATLAS\,9} models agree very well.
\begin{table}
\centering
\caption{Best fitting angular diameters of $\alpha$\,Cet based on
a comparison of our VINCI data to different model atmospheres.}
\label{tab:vinciresults}
\begin{tabular}{rrrrrr}
\hline\hline
\multicolumn{3}{c}{Model parameters}& $\Theta_\mathrm{LD}$ & 
$\Theta_\mathrm{Ross}$ & $\chi^2_\nu$ \\
$T_\mathrm{eff}$ [K]& $\log g$ & $M/$M$_\odot$ & [mas] & [mas] & \\\hline
\multicolumn{6}{l}{{\tt PHOENIX} models:}\\
3800 & 1.0 & 2.3 & 12.60 $\pm$ 0.04 & 12.20 $\pm$ 0.04 & 0.99\\ 
3700 & 1.0 & 2.3 & 12.60 $\pm$ 0.04 & 12.20 $\pm$ 0.04 & 0.98\\ 
3900 & 1.0 & 2.3 & 12.60 $\pm$ 0.04 & 12.20 $\pm$ 0.04 & 0.99\\ 
3800 & 0.5 & 2.3 & 12.95 $\pm$ 0.04 & 12.19 $\pm$ 0.04 & 0.99\\ 
3800 & 1.5 & 2.3 & 12.43 $\pm$ 0.04 & 12.21 $\pm$ 0.04 & 0.99\\ 
3800 & 1.0 & 2.0 & 12.63 $\pm$ 0.04 & 12.20 $\pm$ 0.04 & 0.99\\ 
3800 & 1.0 & 2.6 & 12.58 $\pm$ 0.04 & 12.20 $\pm$ 0.04 & 0.99\\ 
\multicolumn{6}{l}{{\tt ATLAS\,9} model for comparison:}\\
3750 & 1.0 & /   & 12.20 $\pm$ 0.04 & 12.20 $\pm$ 0.04 & 0.99\\\hline
\end{tabular}
\end{table}

The corresponding
synthetic $K$-band visibility curve and thus the resulting
$\chi^2_\nu$ values are not sensitive to variations of $T_\mathrm{eff}$,
$\log g$, and $M$ within the considered ranges. 
The good $\chi^2_\nu$ values of 0.99 shows that the measured and 
model-predicted strength of the limb-darkening effect in the $K$-band
agree well. 
Figure~\ref{fig:vinciplot} shows the measured squared visibility 
amplitudes together with the {\tt PHOENIX} and {\tt ATLAS} 
model prediction. Also shown 
are the simple models of a uniform disc, fully darkened disc, and the
parametrisation $I=\mu^\alpha$ as discussed in Sect.~\ref{sec:vinci}.

The high-precision Rosseland angular diameter obtained from the comparison
of the {\tt PHOENIX} model atmospheres to our VINCI data of 
$\Theta_\mathrm{Ross}$=12.20\,$\pm$\,0.04\,mas corresponds with the 
Hipparcos parallax to a linear Rosseland radius 
of $R_\mathrm{Ross}$=89\,$\pm$\,5\,R$_\odot$. Here, the error is by far
dominated by the error of the bolometric flux. Our VINCI value
for $\Theta_\mathrm{Ross}$ together with the bolometric flux derived above
results in an effective temperature of $T_\mathrm{eff}$=3795\,$\pm$\,70\,K.
$R_\mathrm{Ross}$ and $T_\mathrm{eff}$ result in a luminosity 
$\log L/\mathrm{L}_\odot$=3.16\,$\pm$\,0.08 
($L$=1460\,$L_\odot$\,$\pm$\,300\,$L_\odot$). 
These values are well consistent
but more precise than the values estimated above by a comparison of 
the {\tt PHOENIX} models with available spectrophotometry.
Based on the stellar evolutionary tracks by Girardi et al. (\cite{girardi00}),
we can estimate a mass of $M/M_\odot$=2.3\,$\pm$\,0.2 and a surface gravity 
of $\log g$=0.9\,$\pm$\,0.1 (cgs).
\subsection{Comparison of model predictions to our UVES spectra}
We consider in the following the spherical {\tt PHOENIX} model atmosphere 
as described above with 
parameters $T_\mathrm{eff}$=3800\,K, $\log g$=1.0, and $M=2.3 M_\odot$
because it is closest to the 
model parameters derived by comparison to our VINCI data and well describes
the measured strength of the limb-darkening in the $K$-band. Here, we
compare the model-predicted spectrum in several bandpasses to our measured
UVES spectrum to investigate to what extent the same model that is 
consistent with our interferometric data is consistent with 
our high-resolution spectroscopic data as well.

We use four dominant TiO bandheads
around 5598\AA, 7054\AA, 7088\AA, and 7126\AA for our analysis. 
In addition, we have selected 6 bands around 
dominant atomic lines of Fe\,I (3683\AA, 5447\AA, 6945\AA), Ca\,I (4227\AA), 
and Ti\,I (5966\AA, 9675\AA) that cover all gratings
used for our UVES spectrum and span a total range 
from 360\,nm to 1000\,nm. Profiles of atomic lines and TiO bandheads
such as these are good indicators of stellar model atmosphere parameters,
most importantly, effective temperature and surface gravity (e.g., 
Valenti et al. \cite{valenti98}).
\begin{table}
\centering
\caption{Bandpasses used for the comparison of {\tt PHOENIX} model 
predictions to our UVES spectrum. Listed are the 
central wavelength $\lambda_\mathrm{air}$, the identification (ID), 
the obtained wavelength shift $\Delta\lambda$, the scale factor $f$
between observed and model-predicted spectrum, the reduced $\chi^2_\nu$ 
value, and the ratio of observed and model-predicted equivalent 
width. The scale factors $f$ are derived in blue and red 
bandpasses at $[-8..-4]$\,\AA\ and $[4..8]$\,\AA\ with respect to the 
central wavelength. The $\chi^2_\nu$ values are then derived in 
bandpasses $[-2..2]$\AA\ around $\lambda_\mathrm{air}$.
The equivalent widths are the integrals of the
central bandpasses after normalisation of the spectrum to unity 
using the integral of the blue and red bandpasses.}
\label{tab:uvescomparison}
\begin{tabular}{llrrrr}
\hline\hline
$\lambda_\mathrm{air}$ & ID & $\Delta\lambda$ & $f_\mathrm{norm}$ & 
$\chi^2_\nu$ & $W_\mathrm{obs}/W_\mathrm{model}$\\
\AA                    &    & \AA             &   &              \\\hline
5598.41 & TiO   & -0.031 & 0.89 & 13378 & 0.70 \\        
7054.19 & TiO   & -0.006 & 1.09 &   530 & 1.01 \\        
7087.60 & TiO   & -0.010 & 1.04 &  1071 & 0.80 \\        
7125.59 & TiO   & -0.029 & 1.07 &   674 & 0.91 \\[2ex]   
3683.06 & Fe\,I & -0.002 & 1.09 & 17496 & 3.4  \\        
4226.73 & Ca\,I & -0.004 & 0.68 &   169 & 0.91 \\        
5446.92 & Fe\,I & -0.009 & 0.87 & 32299 & 0.47 \\        
5965.83 & Ti\,I & -0.037 & 0.84 &  7051 & 0.96 \\        
6945.20 & Fe\,I & -0.007 & 1.20 &   718 & 0.69 \\        
9675.49 & Ti\,I & -0.073 & 1.21 &   188 & 0.79 \\\hline  
\end{tabular}
\end{table}
\begin{figure*}
\centering
\resizebox{0.44\hsize}{!}{\includegraphics{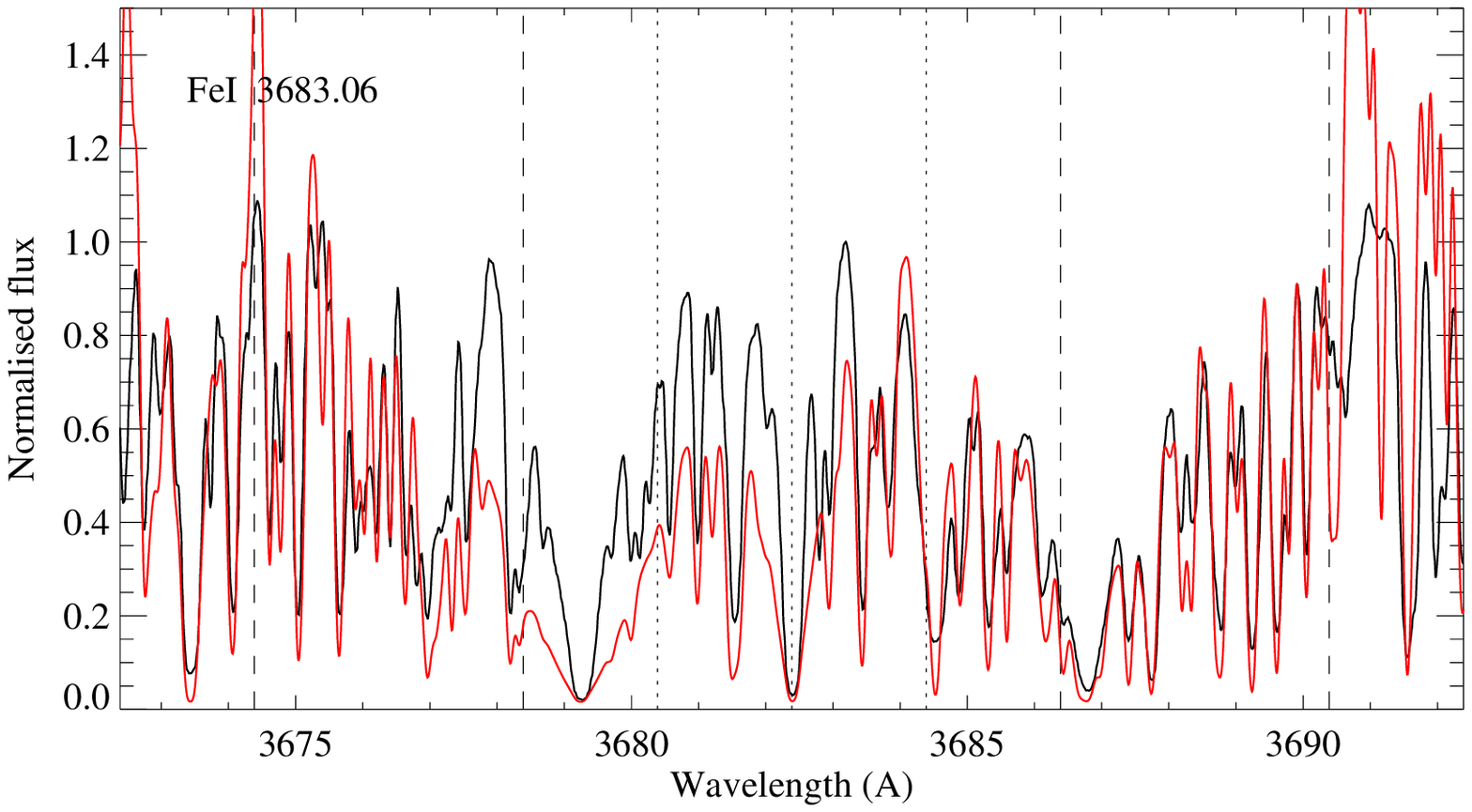}}
\resizebox{0.44\hsize}{!}{\includegraphics{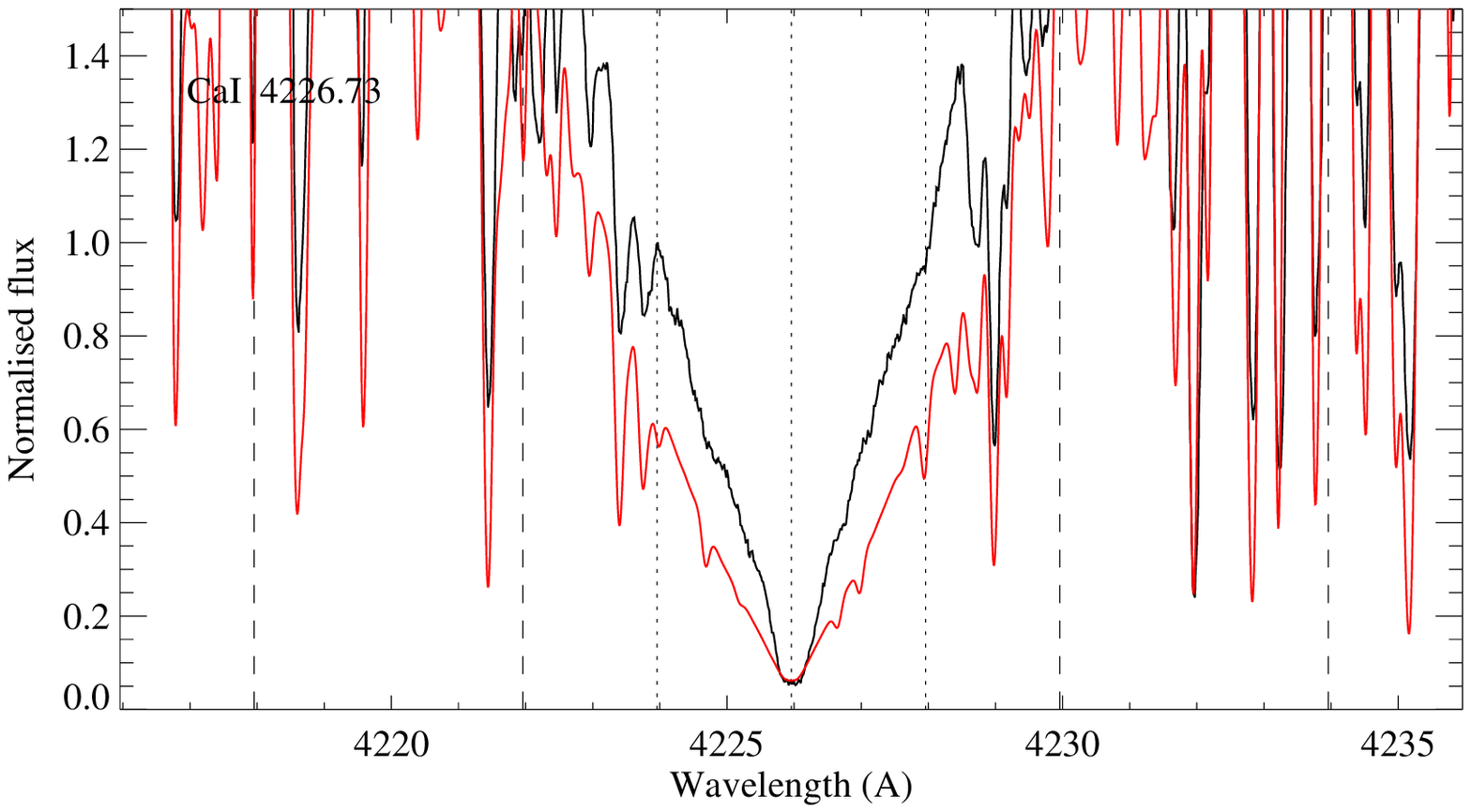}}

\resizebox{0.44\hsize}{!}{\includegraphics{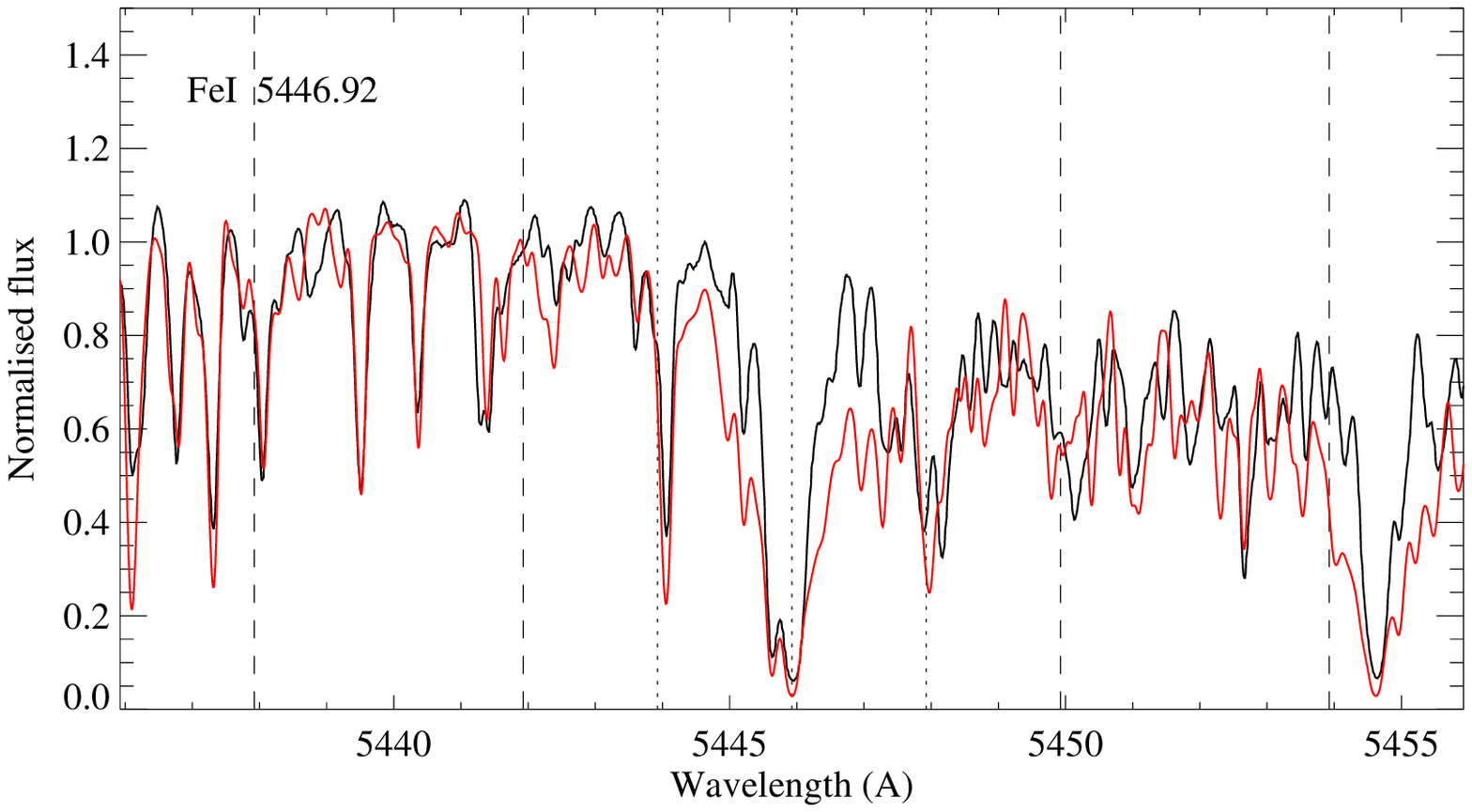}}
\resizebox{0.44\hsize}{!}{\includegraphics{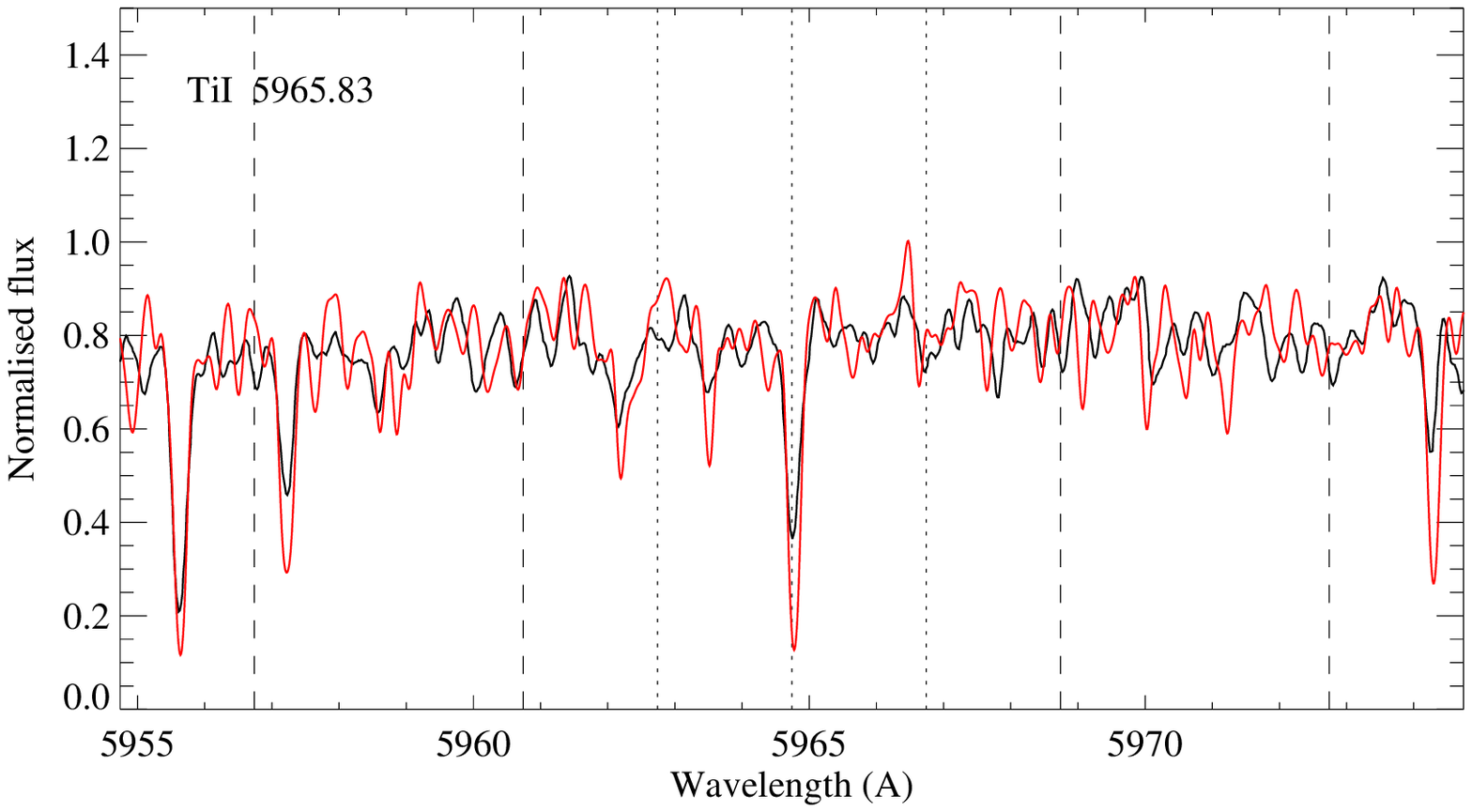}}

\resizebox{0.44\hsize}{!}{\includegraphics{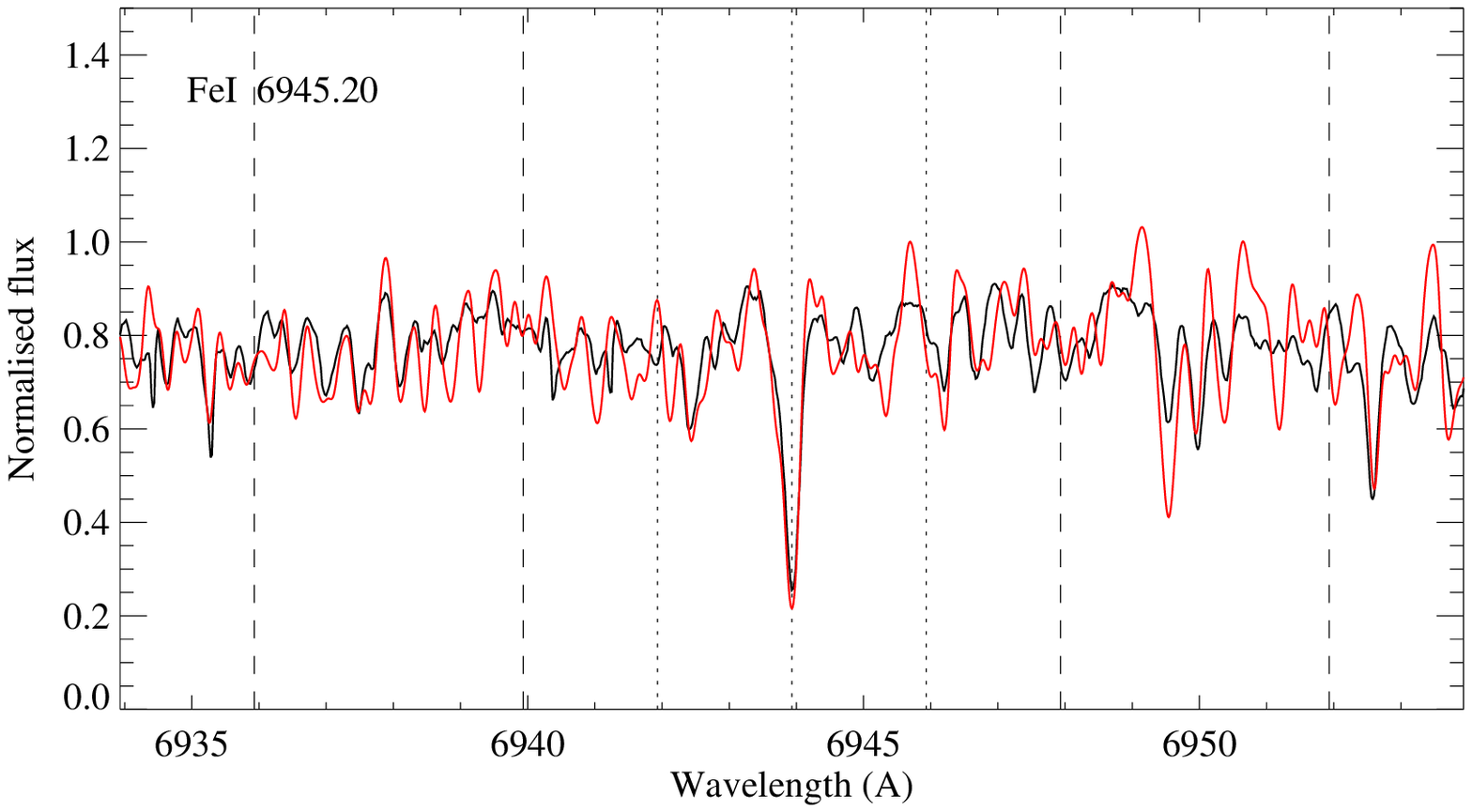}}
\resizebox{0.44\hsize}{!}{\includegraphics{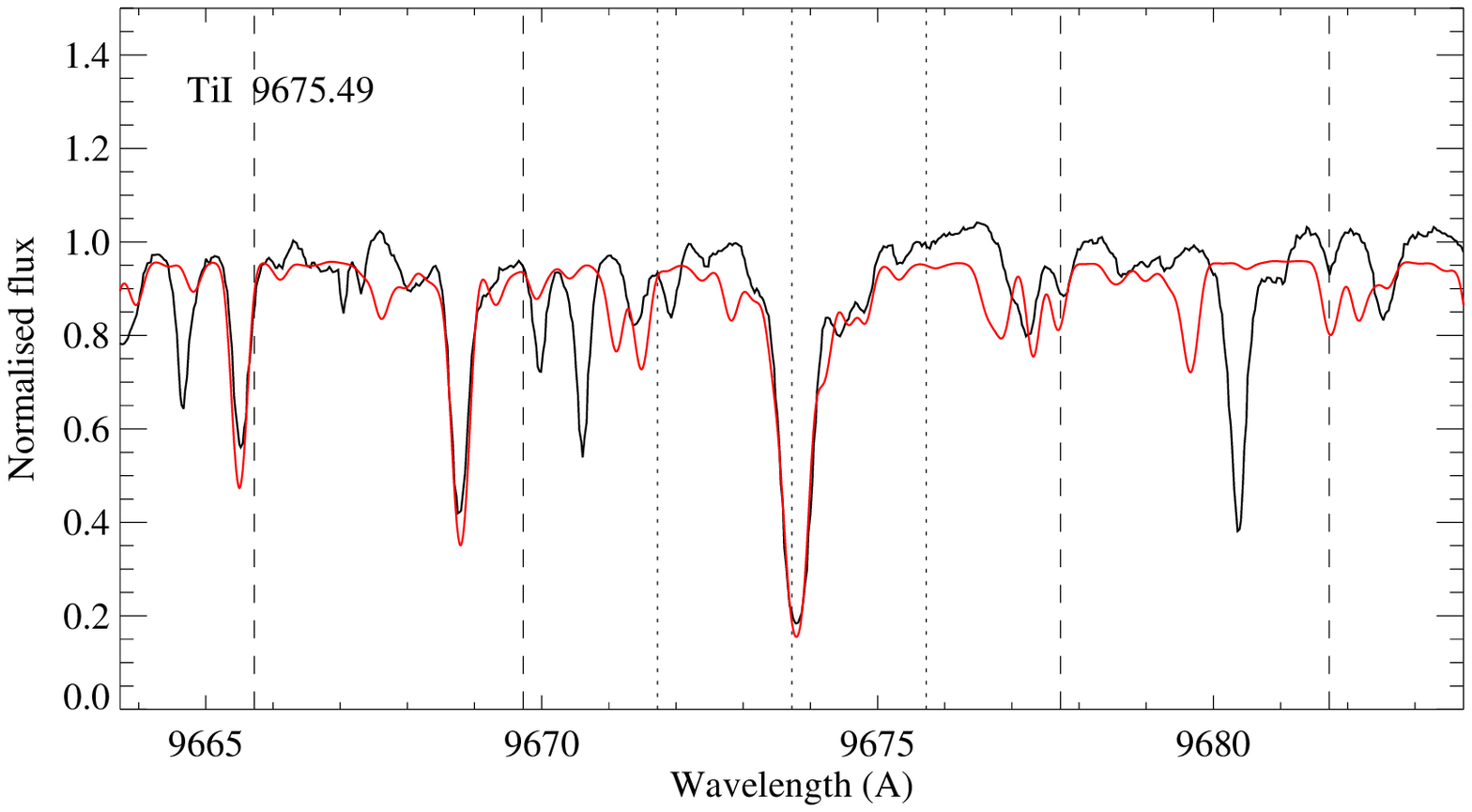}}

\resizebox{0.44\hsize}{!}{\includegraphics{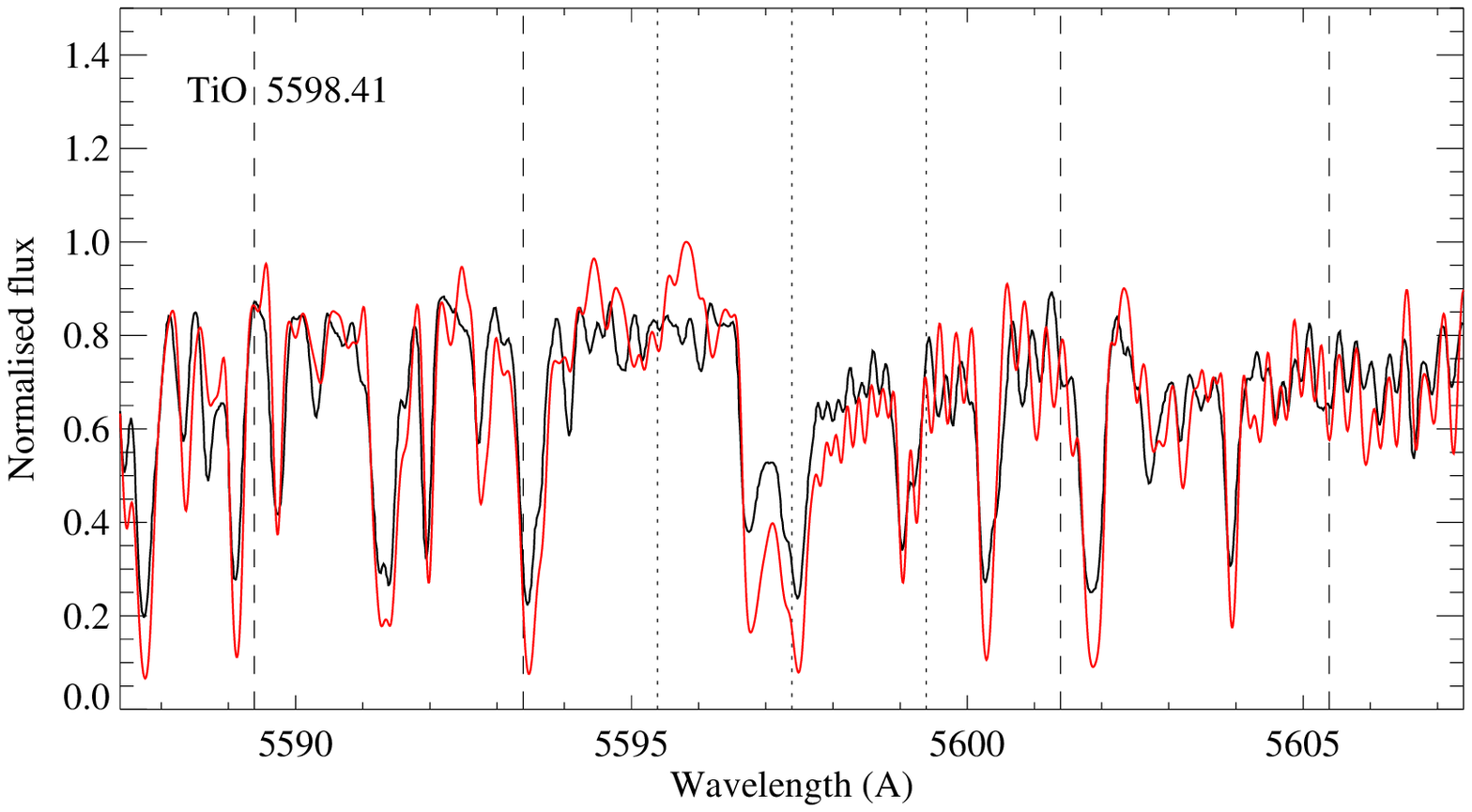}}
\resizebox{0.44\hsize}{!}{\includegraphics{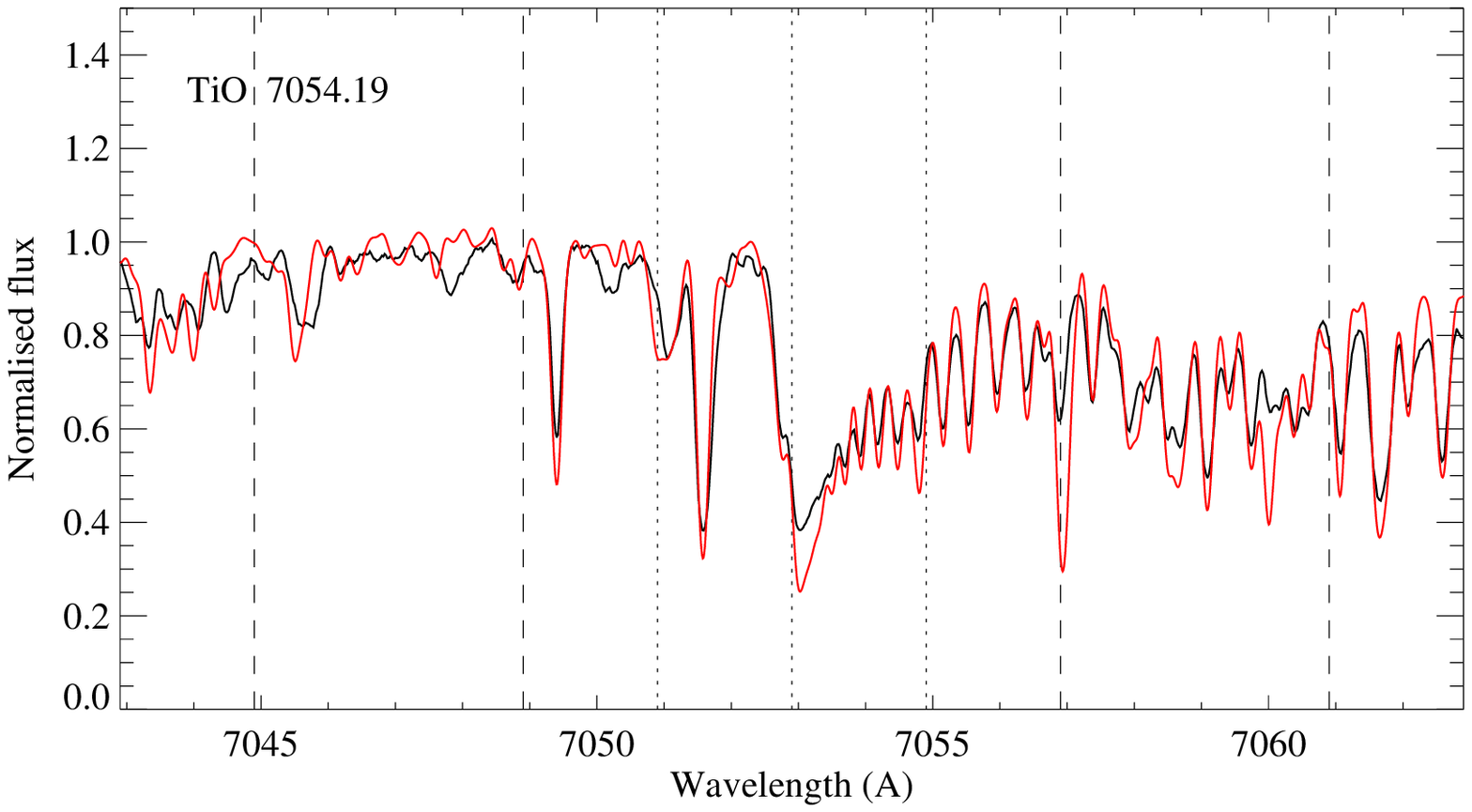}}

\resizebox{0.44\hsize}{!}{\includegraphics{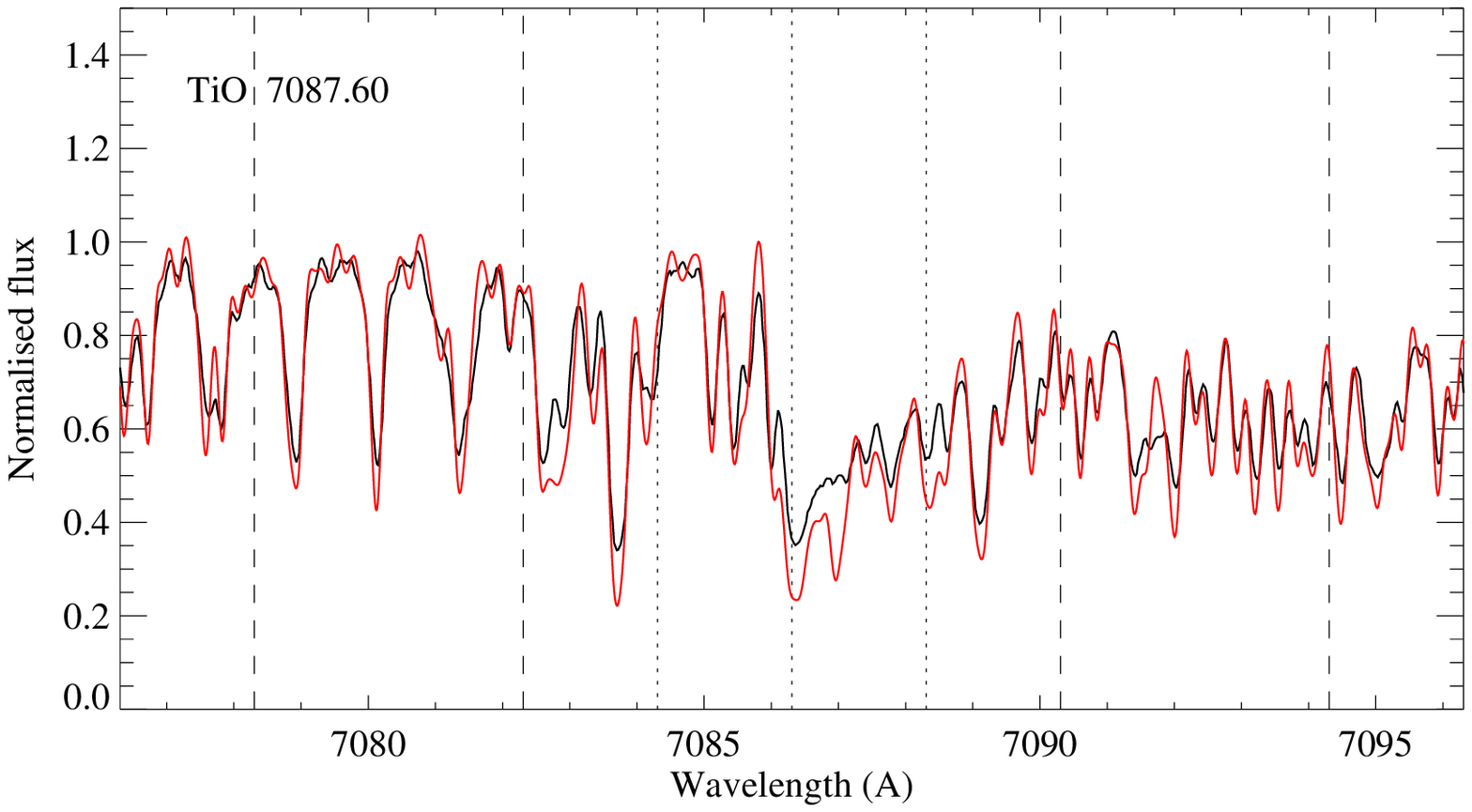}}
\resizebox{0.44\hsize}{!}{\includegraphics{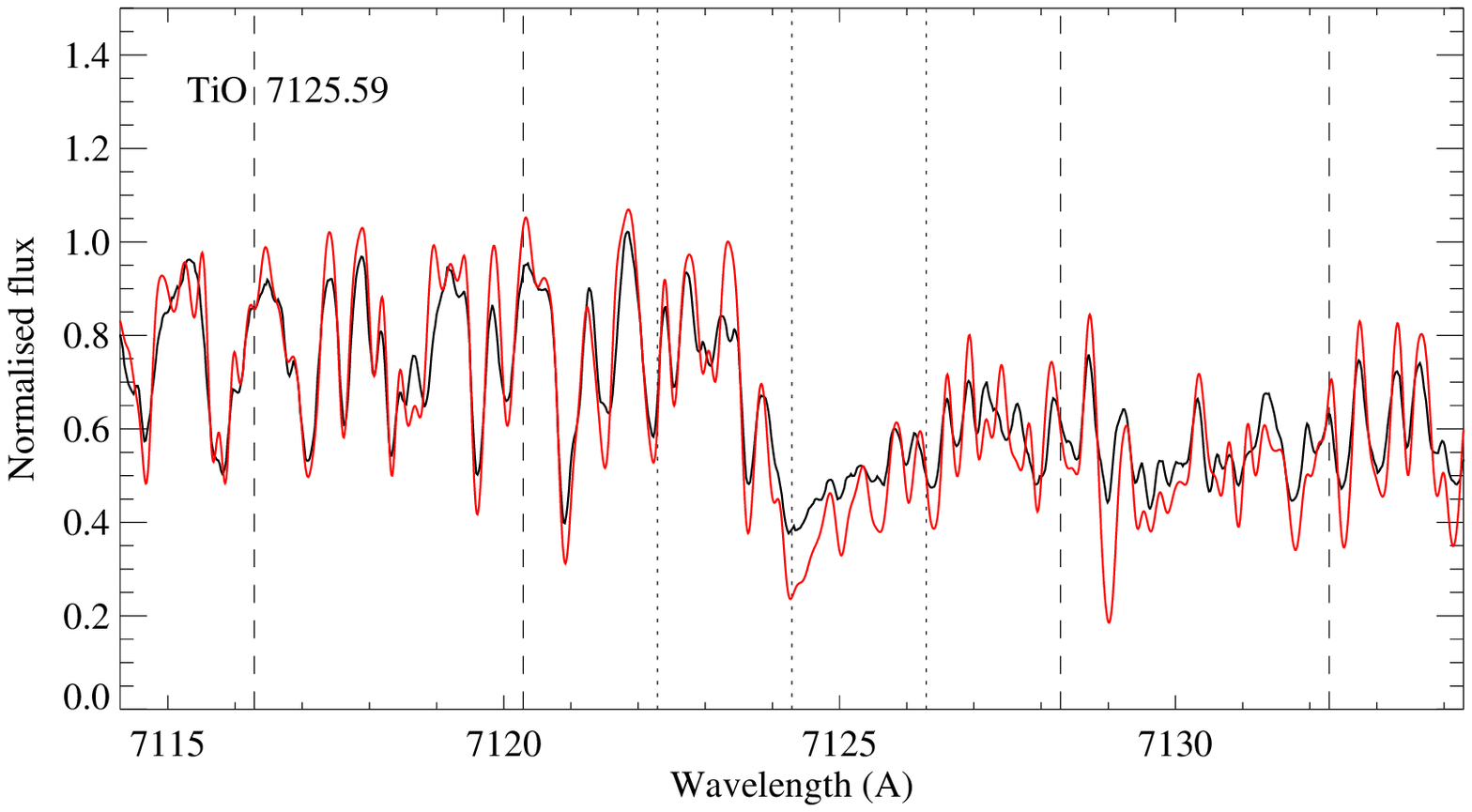}}
\caption{Observed UVES spectrum (black) together with the synthetic 
spectrum (red) for each bandpass considered. The parameters of the
spherical {\tt PHOENIX} model used
are $T_\mathrm{eff}$=3800\,K, $\log g$=1.0, $M=$\,2.3\,$M_\odot$, and
solar chemical abundance.
The vertical dotted 
lines denote the central wavelength and bandpass used
to compute the $\chi^2_\nu$ and integral $W_\mathrm{obs}/W_\mathrm{model}$ 
values in Table~\ref{tab:uvescomparison}.
The vertical dashed lines denote the bandpasses blueward and redward
of the central bandpass that are used to adjust the scale factor $f$
between model spectrum and measured spectrum and to normalise the
equivalent width.}
\label{fig:uves}
\end{figure*}
\paragraph{Calibration of spectral resolution and wavelength}
The wavelength scales of the synthetic spectra 
were corrected for the air refraction index, for the
Earth's motion and rotation, and for the radial velocity
of $\alpha$\,Cet of $v_\mathrm{rad}$=-26.08\,km/sec 
(Famaey et al. \cite{famaey05}).
For each bandpass separately, the synthetic {\tt PHOENIX}
spectrum was broadened by convolution with a rotational profile
using $v \sin i$=3\,km/sec, which is the mean value 
for M\,0 giants (Munari et al. \cite{munari01}).
It was also broadened with a Gaussian profile 
to match the spectral resolution of the observation given 
in Table~\ref{tab:UVES_obs}.
Due to possible residuals of $\alpha$\,Cet's radial velocity, 
and residuals of the synthetic line positions, small offsets between the
synthetic and observed wavelength scales can remain. We have used
the IDL routine {\tt crscor} from the IUE library 
(http://archive.stsci.edu/iue) to cross-correlate the observed and
synthetic spectra for each bandpass used and to derive remaining
wavelength shifts. The resulting wavelength shifts as listed in 
Table~\ref{tab:uvescomparison} are of the order of 0.01\AA\ (corresponding
to 0.6\,km/sec at 5000\AA) and are small for the purpose of our comparison.
\paragraph{Comparison of observed and synthetic spectra}
An absolute flux calibration of our UVES spectra was not obtained,
leading to an unknown scale factor between observed and synthetic spectrum.
We have determined, by a standard least-squares fit,
the best scale factor $f$ to match observed and synthetic spectra
in two bandpasses blueward and redward of each central bandpass.
For a cool giant, we cannot find good continuum bands close to the
central wavelengths.
The bandpasses used for this adjustment have a standard width
of 4\AA\ and standard central wavelengths -6\AA\ and +6\AA\ 
with respect to the central wavelength, and inevitably 
contain spectral features as well.
The resulting $\chi^2_\nu$ value between 
observed and synthetic spectra was calculated for each central 
bandpass. The normalised factors $f_\mathrm{norm}$ 
and the $\chi^2_\nu$ values
are listed in Table~\ref{tab:uvescomparison}. The variations 
of  $f_\mathrm{norm}$ are generally consistent with the relative
uncertainty of the flux calibration of our UVES spectrum of 10-20\%
for each spectrum as quoted in Sect.~\ref{sec:uves}.
In addition to the direct $\chi^2$ comparison between observed and
synthetic spectra, we calculate the ratio between measured and
predicted equivalent width $W_\mathrm{obs}/W_\mathrm{model}$
as a measure of the
consistency of the integrated line strengths. The equivalent width is derived,
separately for synthetic and observed spectra, by integrating 
the spectrum over the central bandpass after
normalisation to unity using the integral of the red and blue bandpasses.

Figure~\ref{fig:uves} shows the final comparison of observed and synthetic
spectra for each considered bandpass. 
Our {\tt PHOENIX} model
describes the measured spectrum around all selected features 
well qualitatively. We have
inspected a number of additional features across the total
wavelength range of our UVES spectrum and find comparable agreement.
However, on the detailed level of the high spectral resolution
obtained, measurement and model prediction exhibit differences that
dominate the resulting $\chi^2_\nu$ value 
in Table~\ref{tab:uvescomparison}. These differences include different
strengths of individual lines, line positions, and features that
appear in the observed spectrum, but not in the synthetic spectrum, and
vice versa. It is known that synthetic and observed stellar spectra at this detailed level show
discrepancies such as these and that relatively large $\chi^2_\nu$ values 
are thus to be expected for a direct comparison 
(see, e.g., Ku{\v c}inskas et al. \cite{kucinskas05}).
The integrated quantity 
$W_\mathrm{obs}/W_\mathrm{model}$ has a mean value of 0.80 and
standard deviation 0.16 (excluding the 3683.06\,\AA\ Fe\,I bandpass where
the blending of different spectral features is strongest), 
i.e., the observed spectral features have, over all, a lower strength 
than predicted by the model at the level of 1.25\,$\sigma$.

A variation of model parameters $\log g$, and $M$
within the uncertainties that remain from the comparison to our
VINCI data do not lead to significant changes of Fig.~\ref{fig:uves}
and on the $\chi^2_\nu$ and $W_\mathrm{obs}/W_\mathrm{model}$ 
values in Table~\ref{fig:uves}. These parameters can thus not be further 
constrained by the comparison to our UVES spectrum.
A variation of $T_\mathrm{eff}$ (model values 3750\,K, 3800\,K, 3850\,K,
3900\,K; other model parameters unchanged) allows us to estimate
the effective temperature of $\alpha$\,Cet 
to $T_\mathrm{eff}\approx$3820\,$\pm\,$50\,K both based on the  
$\chi^2_\nu$ or the $W_\mathrm{obs}/W_\mathrm{model}$ values as a
function of $T_\mathrm{eff}$. This estimate is based on the comparison
of observed and synthetic spectral lines/bands and is consistent 
with the estimate based on the best-fitting diameter derived from the 
interferometric data and the bolometric flux derived from available 
spectrophotometry as described above.

There are three major limitations to a stronger constraint of the model
parameters. Firstly, an absolute flux calibration of the UVES spectrum
was not obtained, and the relative flux calibration within each spectrum
reaches a precision of not better than 10-20\%.
Secondly, for cool giants, the selected spectral features are not isolated
and the bandpasses used are inevitably contaminated by several other lines; 
nearby true continuum bandpasses are not available. Finally, the total 
differences between observed and synthetic spectrum for our bandpasses,
as characterised by the $\chi^2_\nu$ values 
in Table~\ref{tab:uvescomparison}, are dominated by detailed effects
other than the main model parameters $T_\mathrm{eff}$, $\log g$, and $M$.
\subsection{Comparison of model prediction to previous diameter measurements}
\label{sec:litcomp}
\begin{figure}
\centering
\resizebox{1.0\hsize}{!}{\includegraphics{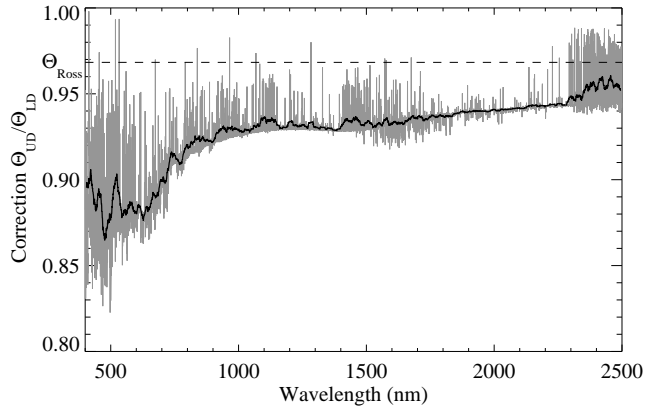}}
\caption{Correction factors from the 0\% LD diameter $\Theta_\mathrm{LD}$
to the UD diameter based on our derived {\tt PHOENIX} model for $\alpha$\,Cet
with parameters $T_\mathrm{eff}$=3800\,K, $\log g$=1, and
$M/M_\odot$=2.3. 
The dashed line denotes the Rosseland diameter $\Theta_\mathrm{Ross}$ with
respect to the 0\% intensity diameter $\Theta_\mathrm{LD}$, which is
0.968 for this model. The thin grey line shows the full resolution of 
our {\tt PHOENIX} model file (5\AA), and the solid black line is reduced
to a resolution of 20\,nm.}
\label{fig:udcorr}
\end{figure}
\begin{table}
\centering
\caption{UD diameters predicted for several previously used
bandpasses based on our derived {\tt PHOENIX} model for $\alpha$\,Cet 
with parameters $T_\mathrm{eff}$=3800\,K, $\log g$=1,
$M/M_\odot$=2.3, solar chemical abundance, and $\Theta_\mathrm{LD}$=12.60\,mas.
We estimate the accuracy of the model predictions to be $\le 5\%$, see
text for details on the calculation.}
\label{tab:udcorr}
\begin{tabular}{rrrrl}
\hline\hline
$\lambda_c$ & $\delta\lambda$ & 
$\Theta_\mathrm{UD}^\mathrm{Model}$ &
$\Theta_\mathrm{UD}^\mathrm{Lit.}$  & Reference \\
nm & nm & mas & mas & \\\hline
2187 & 400 & 11.90 & 11.95 $\pm$ 0.06 & Direct fit in this work\\[1ex]
450  & 20  & 11.24 & 11.33 $\pm$ 0.29 & Mozurkewich et al. (\cite{mozurkewich91})\\
451  & 20  & 11.24 & 11.33 $\pm$ 0.41 & Mozurkewich et al. (\cite{mozurkewich03})\\
550  & 20  & 11.09 & 11.47 $\pm$ 0.25 & Mozurkewich et al. (\cite{mozurkewich03})\\
712  & 12  & 11.31 & 11.95 $\pm$ 0.23 & Quirrenbach et al. (\cite{quirrenbach93})\\
754  & 5   & 11.52 & 11.66 $\pm$ 0.22 & Quirrenbach et al. (\cite{quirrenbach93})\\
800  & 20  & 11.60 & 12.25 $\pm$ 0.16 & Mozurkewich et al. (\cite{mozurkewich91})\\
800  & 20  & 11.60 & 12.27 $\pm$ 0.24 & Mozurkewich et al. (\cite{mozurkewich03})\\
2200 & 400 & 11.90 & 11.6  $\pm$ 0.4  & Dyck et al. (\cite{dyck98})\\\hline
\end{tabular}
\end{table}
Several UD diameter measurements of $\alpha$\,Cet based on long-baseline
interferometry have previously been obtained (see Sect.~\ref{sec:intro}). 
Here, we use our derived {\tt PHOENIX} model 
of $\alpha$\,Cet ($T_\mathrm{eff}$=3800\,K, $\log g$=1,
$M/M_\odot$=2.3, solar chemical abundance, $\Theta_\mathrm{LD}$=12.60\,mas) 
to predict the UD diameter at the previously used bandpasses. 
The correction factor from the 0\% (LD) diameter to the UD diameter
is calculated so that both visibility curves, the UD curve and the
curve based on the model atmosphere, match where $|V|^2=0.3$, as
previously done by, e.g., Hanbury Brown et al. (\cite{hanbury74}), 
Quirrenbach et al. (\cite{quirrenbach96}), and
Wittkowski et al. (\cite{wittkowski01}). The choice to match the 
visibility curves at $|V|^2=0.3$ is arbitrary, and the accuracy of the
resulting correction factor may decrease if most measured visibility data 
were obtained at other parts of the visibility curve. We average the 
correction factors over rectangular bandpasses with central wavelength
$\lambda_c$ and width $\delta\lambda$. The exact shape of the 
bandpasses used for the previous observations and the exact signal processing 
of the different instruments is not taken into account. We estimate the total
error resulting from these simplifications to be less than about 5\%.
Figure~\ref{fig:udcorr} shows the resulting correction factors for the 
wavelength range from 400\,nm to 2500\,nm. 
Table~\ref{tab:udcorr} lists the previously obtained interferometric
UD diameter measurements of $\alpha$\,Cet compared to the prediction by
our derived {\tt PHOENIX} model for each specific bandpass used.
Only the measurements at 550\,nm and 800\,nm by Mozurkewich et al. 
(\cite{mozurkewich91,mozurkewich03}) and the measurement 
at 712\,nm (TiO band) by Quirrenbach et al. (\cite{quirrenbach93})
are not consistent within 1\,$\sigma$ with the nominal value
of our model prediction, but still within the estimated $\sim$\,5\% 
uncertainty of our prediction. The discrepancy at
the TiO bandpass may either be caused by an imperfect match of the exact
instrumental bandpass and signal processing (as here the strength of the 
limb-darkening effect changes rapidly), or by an imperfect model description
of the spatial structure of the layers where TiO molecules reside.
It has been reported that the use of different line list
combinations of molecules lead to significantly different model
structures and spectra, in particular in the optical where TiO bands are
important (Allard et al. \cite{allard00}; 
Ku{\v c}inskas et al. \cite{kucinskas05}).
\section{Summary and discussion}
\begin{table}
\centering
\caption{Fundamental parameters of the M\,1.5 giant $\alpha$\,Cet
based on the analysis in this paper. The mass and surface gravity
rely on the evolutionary tracks by Girardi et al. (\cite{girardi00}).}
\label{tab:fundpar}
\begin{tabular}{ll}
\hline\hline
Parameter & Value \\\hline
Rosseland angular diameter & $\Theta_\mathrm{Ross}=12.20\pm 0.04$\,mas\\
Rosseland linear radius  & $R_\mathrm{Ross}=89\pm 5 R_\odot$\\
Bolometric flux & $f_\mathrm{bol}=(1.03\pm 0.07)\times 10^{-12}$\,W/m$^2$\\
Effective temperature & $T_\mathrm{eff}=3795\pm 70$\,K\\
Luminosity & $\log L/L_\odot=3.16\pm 0.08$\\
Mass & $M=2.3\pm 0.2 M_\odot$\\
Surface gravity & $\log g=0.9\pm 0.1$ (cgs)\\\hline
\end{tabular}
\end{table}
\begin{figure}
\centering
\resizebox{1.0\hsize}{!}{\includegraphics{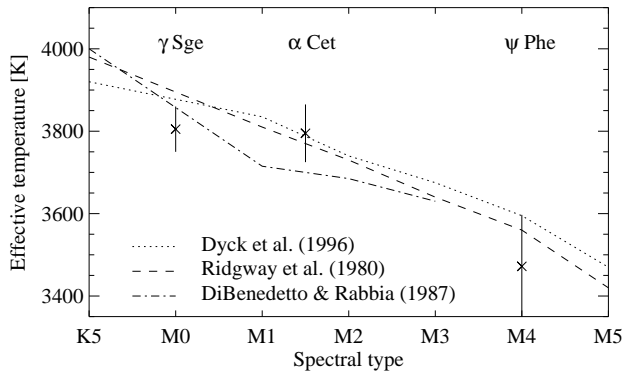}}
\caption{Measured effective temperature of the
M giants $\gamma$\,Sge (Wittkowski et al. \cite{wittkowski06}),
$\alpha$\,Cet (present paper), 
and $\psi$\,Phe (Wittkowski et al. \cite{wittkowski04}) versus spectral
type compared to empirical calibrations for giants. The three measurements 
are consistently obtained by Rosseland diameters based on a 
comparison of VINCI interferometric data to spherical {\tt PHOENIX} 
model atmospheres and bolometric fluxes
obtained by careful integration of available spectro-photometry.}
\label{fig:giants}
\end{figure}
We have shown that our derived {\tt PHOENIX} model atmosphere for
$\alpha$\,Cet (Menkar) with model parameters 
$T_\mathrm{eff}$=3800\,K, $\log g$=1, and $M/M_\odot$=2.3
is consistent with both the measured strength
of the limb-darkening in the broad near-infrared $K$-band and with the 
profiles of spectral bands around selected atomic lines 
and TiO bandheads from 370\,nm to 1000\,nm. At the detailed level
of our high spectral resolution ($R$ up to 110\,000), however, noticeable
differences of observed and synthetic spectra exist. The discrepancies
include differences in the strengths and positions of spectral lines/bands,
as well as detailed spectral features that only appear in either the observed 
or synthetic spectrum. It has previously been reported that the existence
of such detailed effects, especially in the optical where TiO bands are 
important, is known and may be due to, for instance, different line list 
combinations or treatments of convection (Allard et al. \cite{allard00}, 
Ku{\v c}inskas et al. \cite{kucinskas05}).

The fundamental stellar parameters of $\alpha$\,Cet are most 
constrained by our high-precision angular diameter obtained from the
comparison of the {\tt PHOENIX} atmosphere model with our VINCI data,
and the bolometric flux based on available spectro-photometry.
This resulting set of fundamental stellar parameters of $\alpha$\,Cet
is summarised in Table~\ref{tab:fundpar}.
These values are consistent with previous measurements mentioned in 
Sects.~\ref{sec:intro} and \ref{sec:litcomp}, 
but generally have a higher precision.

Together with the results on the M\,4 giant $\psi$\,Phe (Wittkowski et al.
\cite{wittkowski04}, Paper\,II of this series), 
the M\,0 giant $\gamma$\,Sge (Wittkowski et al. 
\cite{wittkowski06}, Paper\,III of this series), 
and the M\,1.5 giant $\alpha$\,Cet (present paper),
we have consistently obtained sets of high-precision fundamental parameters
for these three M giants based on a comparison of VINCI data to 
spherical {\tt PHOENIX} model atmospheres. Figure~\ref{fig:giants} shows
the obtained effective temperatures versus spectral types compared to
empirical calibrations by Ridgway et al. (\cite{ridgway80}),
Di Benedetto \& Rabbia (\cite{benedetto87}), and Dyck et al. (\cite{dyck96}).
The measurements are consistent
with the empirical calibrations shown at about the 1\,$\sigma$ level,
but may suggest a flatter temperature slope for early M giants and a stronger
decrease of effective temperature for cooler (toward M\,4-5) giants. 

The studies on these three cool giants also consistently show that the model-predicted strength of the 
limb-darkening is not significantly
affected by the geometry of the model atmosphere for the broad 
near-infrared $K$-band, and also for narrower visual bandpasses 
in the case of the NPOI observations of $\gamma$\,Sagittae of Paper\,III, 
with the precision of current measurements. 
Effects from plane-parallel versus spherical geometry
thus appear to be more subtle than generally seems to be expected. 
However, the most precise definition of a meaningful diameter, such as 
the diameter where the Rosseland-mean optical depth reaches unity, can 
best be obtained based on a spherical model, as discussed in the 
articles of this series.

Our results illustrate as well the power of combining interferometry 
and high-resolution spectroscopy to constrain and calibrate stellar model 
atmospheres.
The newly offered instrument AMBER (Petrov et al. \cite{petrov03})
at the VLTI permits the recording of spectro-interferometric 
data with a spectral resolution of up to 12\,000, i.e., with a resolution that
is several orders of magnitudes above that of previous interferometric
instruments.
This spectral resolution of the newest generation of interferometric 
instruments, however, is still clearly below that of modern 
optical spectrographs, such as UVES, or IR spectrographs, such as CRIRES, 
which provide a spectral resolution of up to $R\ge 100\,000$.
It will thus be very valuable to combine AMBER interferometry 
with UVES or CRIRES highest resolution spectroscopy to constrain stellar
models. A limitation of our current study is the lack of an absolute
flux calibration of our UVES data, which should be obtained with high
precision for such suggested future studies.
\begin{acknowledgements}
This work was performed in part (JPA) under contract with the Jet Propulsion
Laboratory (JPL) funded by NASA through the Michelson Fellowship Program.
JPL is managed for NASA by the California Institute of Technology.
\end{acknowledgements}
{}
\end{document}